\documentclass[smallextended]{svjour3}       
\smartqed  
\usepackage{graphicx,epsfig}

\begin{document}

\title{
Phase diagram for the one-way quantum deficit of two-qubit X states
}

\author{
M.A.Yurischev
}

\institute{
M.~A.~Yurischev
\at
Institute of Problems of Chemical Physics, Russian Academy of Sciences,
Chernogolovka 142432, Moscow Region, Russia\\     
\email{yur@itp.ac.ru} }

\date{Received:}

\maketitle

\begin{abstract}
The one-way quantum deficit, a measure of quantum correlation, can exhibit for X
quantum states the regions (subdomains) with the phases $\Delta_0$ and
$\Delta_{\pi/2}$ which are characterized by {\em constant\/} (i.e., universal)
optimal measurement angles, correspondingly, zero and $\pi/2$ with respect to
the $z$-axis and a third phase $\Delta_\vartheta$ with the {\em variable\/}
(state-dependent) optimal measurement angle $\vartheta$.
We build the complete phase diagram of one-way quantum deficit for the
XXZ subclass of symmetric X states.
In contrast to the quantum discord where the region for the phase with variable
optimal measurement angle is very tiny (more exactly, it is a very thin layer),
the similar region $\Delta_\vartheta$ is large and achieves the sizes comparable
to those of regions $\Delta_0$ and $\Delta_{\pi/2}$.
This instils hope to detect the mysterious fraction of quantum correlation with
the variable optimal measurement angle experimentally.
\end{abstract}

%

\section{Introduction}
\label{sect:Intro}
Quantum correlations play the key role in quantum information science.
Many kinds of quantum correlations have been introduced so far
and now their properties are scrupulously analyzed.
One of the most important places among such correlations
beyond quantum entanglement belongs to the quantum discord and
one-way quantum work (information) deficit
\cite{MBCPV12,Str15,ABC16,FPA17,BDSRSS18,OHHH02,HHHHOSS02,HHHOSSS05,Z03}.
In the present paper we focus on the latter measure of quantum correlation.
The one-way deficit has operational interpretation in thermodynamics and is equal
to a slightly different version of quantum discord \cite{Z03} called in
Ref.~\cite{MBCPV12} as the ``thermal discord'';
this term is also supported in the topical review~\cite{ABC16}.

Remarkably, both varieties of discord definition yield the same results for the
Bell-diagonal states and even for the X quantum states if one qubit of the system
is maximally mixed and the measurements are performed on this qubit \cite{YF16}.
However, in spite of closeness of definitions, the quantum discord and one-way
quantum deficit may lead, generally speaking, to the quite different quantitative
and actually qualitative behavior of quantum correlation in more general cases.

It is known \cite{CRC10,VR12,H13,YWF16} that the optimization of quantum discord
as well as one-way deficit for X states is reduced to a minimization only on one
variable, namely the polar angle $\theta\in[0,\pi/2]$ (the angle of deviation from the
$z$-axis of X state) whereas the azimuthal angle $\phi$ can be always eliminated by
local rotations around the $z$-axis.
A minimization procedure for measurement-dependent discord, $Q(\theta)$, and one-way
deficit, $\Delta(\theta)$, leads to the optimized functions of discord
$Q=\min\{Q_0,Q_{\pi/2},Q_{\theta^*}\}$ (subscripts 0, $\pi/2$, and $\theta^*$ denoting
the corresponding optimal measured angles for the quantum discord) and one-way deficit
${\rm\Delta}=\min\{\Delta_0,\Delta_{\pi/2},\Delta_\vartheta\}$ in their domain of
definition are the piecewise-defined ones.
In other words, the total domain of definition consists of subdomains each one
corresponding to the own branch (phase or fraction - in physical language).
Important problem is to separate all possible phases of quantum correlation
and find in fact the exhausted phase diagram.

Recently \cite{Y17} the quantum discord for the symmetric XXZ subclass of X states
has been considered and the three-dimensional phase diagram for it has been obtained.
In this paper we perform calculations for the same subclass but this time for the
one-way quantum deficit.
This allows us to compare the behavior of both measures of quantum correlation and
reveal differences between them.
One of the most surprising results is that the ``anomalous'' subdomain with variable
optimal measurement angle is essentially larger than that for the quantum discord
and its sizes achieve the values which make it accessible for experimental
investigations.

The reminder of this paper is organized as follows.
In the next section we discuss the possibility to create the complete phase diagram
for the general two-qubit X quantum state.
The domain of definition of one-way deficit function for the symmetric XXZ quantum
state is obtained in Sec.~3.
All necessary formulas for the branches of piecewise one-way deficit function
are presented in Sec.~4.
Equations for the boundaries between different fractions of one-way deficit are given
in Sec.~5.
In Sec.~6 we solve the equations for the boundaries and discuss the results obtained.
Concluding remarks and possible perspectives are summarized in Sec.~7.

\section{
General X state and atlas of phase diagrams 
}
\label{sect:genX}
In the most general case, the X quantum state of two-qubit system $AB$ can be
written as
\begin{eqnarray}
   \label{eq:rho7-Bloch}
   \rho_{AB}&=&4^{-1}(1
   + s_1\sigma_z\otimes1
   + s_21\otimes\sigma_z
   + c_1\sigma_x\otimes\sigma_x 
   + c_2\sigma_y\otimes\sigma_y 
   + c_{12}\sigma_x\otimes\sigma_y 
	 \nonumber\\
   &+& c_{21}\sigma_y\otimes\sigma_x 
   + c_3\sigma_z\otimes\sigma_z),
\end{eqnarray}
where $\sigma_\alpha$ ($\alpha=x,y,z$) is the vector of the Pauli matrices.
The density matrix (\ref{eq:rho7-Bloch}) contains seven real parameters
$s_1$, $s_2$, $c_1$, $c_2$, $c_{12}$, $c_{21}$, and $c_3$
which are the unary and binary correlation functions, i.e., experimentally
measurable quantities:
\begin{eqnarray}
   \label{eq:corr-s1c3}
   &&\qquad
	 s_1=\langle\sigma^1_z\rangle={\rm Tr}(\rho_{AB}\sigma_z\otimes1),\quad
   s_2=\langle\sigma^2_z\rangle={\rm Tr}(\rho_{AB}1\otimes\sigma_z),
   \nonumber\\
   &&\
	 c_1=\langle\sigma^1_x\sigma^2_x\rangle={\rm Tr}(\rho_{AB}\sigma_x\otimes\sigma_x),\quad
   c_2=\langle\sigma^1_y\sigma^2_y\rangle={\rm Tr}(\rho_{AB}\sigma_y\otimes\sigma_y),
   \nonumber\\
   &&c_{12}=\langle\sigma^1_x\sigma^2_y\rangle={\rm Tr}(\rho_{AB}\sigma_x\otimes\sigma_y),\quad
   c_{21}=\langle\sigma^1_y\sigma^2_x\rangle={\rm Tr}(\rho_{AB}\sigma_y\otimes\sigma_x),\\
   &&\qquad\qquad\qquad\qquad
	 c_3=\langle\sigma^1_z\sigma^2_z\rangle={\rm Tr}(\rho_{AB}\sigma_z\otimes\sigma_z).
   \nonumber
\end{eqnarray}
It is clear that
\begin{equation}
   \label{eq:s1c3}
   -1\le s_1, s_2, c_1, c_2, c_{12}, c_{21}, c_3\le 1,
\end{equation}
hence in any case the parameters do not come out of the seven-dimensional cube
in the space R$\!^7$.

The density matrix $\rho_{AB}$ in open form reads
\begin{equation}
   \label{eq:rho7}
   \rho_{AB}
	 =\frac{1}{4}\!\left(
      \begin{array}{cccc}
      1+s_1+s_2+c_3&0&0&c_1-c_2\\
      &&&-i(c_{12}+c_{21})\\
      0&1+s_1-s_2-c_3&c_1+c_2&0\\
      &&+i(c_{12}-c_{21})&\\
      0&c_1+c_2&1-s_1+s_2-c_3&0\\
      &-i(c_{12}-c_{21})&&\\
      c_1-c_2&0&0&1-s_1-s_2+c_3\\
      +i(c_{12}+c_{21})&&&
      \end{array}
   \right)
\end{equation}
that demonstrates the obvious X structure.

Quantum correlations are invariant under the local unitary transformations
of density matrix \cite{BM12}.
Thanks to this fundamental property one can eliminate the complex phases in
the off-diagonal elements of density matrix (\ref{eq:rho7}) and reduce it
with the help of local rotations around the $z$-axis,
$\exp(-i\varphi_1\sigma_z/2)\otimes\exp(-i\varphi_2\sigma_z/2)$,
to the {\em real} X form (see \cite{MBCPV12} and, e.g., \cite{Y14,Y14a}):
\begin{equation}
   \label{eq:rho5}
   \rho_{AB}
	 =\frac{1}{4}\!\left(
      \begin{array}{cccc}
      1+s_1+s_2+c_3&0&0&u\\
      0&1+s_1-s_2-c_3&v&0\\
      0&v&1-s_1+s_2-c_3&0\\
      u&0&0&1-s_1-s_2+c_3\\ 
      \end{array}
   \right)
\end{equation}
with
\begin{equation}
   \label{eq:uv}
   u=[(c_1-c_2)^2+(c_{12}+c_{21})^2]^{1/2},\qquad
   v=[(c_1+c_2)^2+(c_{12}-c_{21})^2]^{1/2}.
\end{equation}
This matrix contains five real parameters: $s_1$, $s_2$, $c_3$, $u$, and $v$.
All kinds of quantum correlations in the state (\ref{eq:rho5}) are the same as
in the original seven-parameter state (\ref{eq:rho7-Bloch}).
Moreover, since off-diagonal entries are now non-negative,
this will allow us to set by optimization the azimuthal angle $\phi=0$
\cite{CRC10,H13}.
(In fact, this nonnegativity is enough only for the product of off-diagonal
entries, $uv\geq0$.)

Return to the original density matrix (\ref{eq:rho7}).
Due to the  requirement of positive semidefiniteness for any density matrix,
one obtains restrictions for the model parameters
\begin{eqnarray}
   \label{eq:7D}
   (1+c_3)^2-(s_1+s_2)^2\geq(c_1-c_2)^2+(c_{12}+c_{21})^2=u^2,\nonumber\\
   (1-c_3)^2-(s_1-s_2)^2\geq(c_1+c_2)^2+(c_{12}-c_{21})^2=v^2.
\end{eqnarray}
A body which is bounded by these two intersecting surfaces of second order
is the domain of definition for the arguments of quantum correlation functions.
Note that this domain lies in the seven-dimensional hypercube $[-1,1]^{\times7}$.

Classification and construction of all possible phases of quantum correlation in the
whole seven-parameters space is an enormous task which as yet only waits for
its solution.
One way to solve it is to build an atlas of maps \cite{P87}, i.e.,
the collection of two- or three-dimensional phase diagrams.
We now pass on to the first, likely simplest but important ``map''
of such an atlas.

\section{
Density matrix for the symmetric XXZ model and domain of definition
}
\label{sect:DensMatr}
As for the quantum discord \cite{Y17}, in this paper we restrict ourselves by
the same three-dimensional space, i.e., set $s_1=s_2$ (the mirror symmetry
$C_s$ of two-qubit system),
$c_{12}=c_{21}=0$, and $c_1=c_2$ [the axial spin symmetry $U(1)$].
In this case the density matrix is written as
\begin{equation}
   \label{eq:rhoXXZS-Bloch}
   \rho_{AB}=4^{-1}[1
   + s_1(\sigma_z\otimes1
   + 1\otimes\sigma_z)
   + c_1(\sigma_x\otimes\sigma_x 
   + \sigma_y\otimes\sigma_y) 
   + c_3\sigma_z\otimes\sigma_z].
\end{equation}
We will call this state as the symmetric XXZ one ---
by analogy with the well-known statistical-mechanical Hamiltonians \cite{B82,F14}.
Notice that the $z$-component of total spin,
$S_z=(\sigma_z\otimes1+1\otimes\sigma_z)/2$,
commutes with the density matrix (\ref{eq:rhoXXZS-Bloch});
this is a sequence of inner symmetry
$U(1)=\{\exp(i\varphi S_z)|\,\varphi\in[0,2\pi]\}$.
As a result, the full symmetry group is $C_s\times U(1)$.

In the open form the density matrix (\ref{eq:rhoXXZS-Bloch}) is given by
\begin{equation}
   \label{eq:rhoXXZS}
   \rho_{AB}
	 =\frac{1}{4}\!\left(
      \begin{array}{cccc}
      1+2s_1+c_3&0&0&0\\
      0&1-c_3&2c_1&0\\
      0&2c_1&1-c_3&0\\
      0&0&0&1-2s_1+c_3
      \end{array}
   \right).
\end{equation}
Three-parameter quantum states with such a block-diagonal structure are
discussed in connection with the problem of maximally entangled mixed states (MEMS)
\cite{BMNM04} (see also \cite{MMH17}).

\begin{figure}[t]
\begin{center}
\epsfig{file=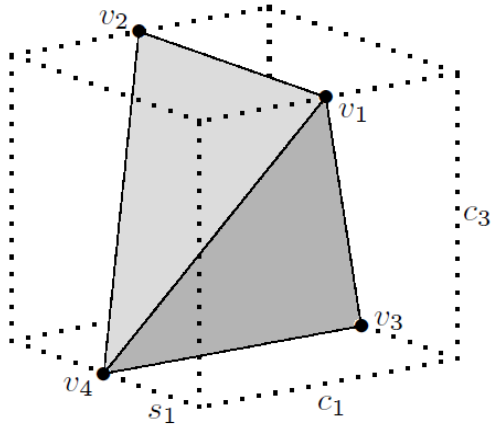,width=7cm}
\caption{
Shaded tetrahedron ${\cal T}$ embedded in a three-dimensional cube
(dotted lines) is the domain of definition for the parameters (arguments)
$s_1$, $c_1$, and $c_3$
}
\label{fig:z_xxz-m2}
\end{center}
\end{figure}

Restrictions (\ref{eq:7D}) are reduced here to the conditions
\begin{equation}
   \label{eq:c3s1c1}
   c_3\in[-1,1],\quad
   s_1\in[-(1+c_3)/2,(1+c_3)/2],\quad
	 c_1\in[-(1-c_3)/2,(1-c_3)/2]
\end{equation}
which define a tetrahedron ${\cal T}$ (see Fig.~\ref{fig:z_xxz-m2}).
This tetrahedron is enclosed in the three-dimensional cube $[-1,1]^{\otimes3}$,
has the vertices $v_1$, $v_2$, $v_3$, and $v_4$ and
isosceles triangle faces.
Volume of tetrahedron ${\cal T}$ equals one sixth part of cube volume ($=2^3$).
So, one may say that the one-way deficit ${\rm\Delta}(s_1,c_1,c_3)$ is a function
on the tetrahedron ${\cal T}$.

Notice that the symmetric XXZ states (\ref{eq:rhoXXZS-Bloch}) may be written
in an equivalent form which is important for the MEMS problem \cite{IH00,HI00},
\begin{equation}
   \label{eq:rho_q1-4}
   \rho_{AB}=q_1|\Psi^+\rangle\langle\Psi^+| + q_2|\Psi^-\rangle\langle\Psi^-|
	 + q_3|00\rangle\langle 00|
	 + q_4|11\rangle\langle 11|,
\end{equation}
where $q_1+q_2+q_3+q_4=1$,
\begin{equation}
   \label{eq:q1-4}
   q_{1,2}=(1\pm2c_1-c_3)/4,\qquad q_{3,4}=(1\pm2s_1+c_3)/4,
\end{equation}
$|\Psi^{\pm}\rangle=(|01\rangle\pm|10\rangle)/\sqrt{2}$ are the Bell states, and
$|00\rangle$ and $|11\rangle$ are product-states orthogonal to $|\Psi^{\pm}\rangle$;
they represent two ``up'' and ``down'' oriented spins (or two horizontally and
vertically polarized photons), respectively.
One may exchange $|\Psi^+\rangle\leftrightarrow|\Psi^-\rangle$ and
$|00\rangle\leftrightarrow|11\rangle$ because they belong to the local unitary
transformations and therefore do not change the value of quantum correlation.
Quantities $q_i~(i=1,\cdots,4)$ are equal to the eigenvalues of density matrix and
therefore must be non-negative.
Hence the domain of definition in the space $(q_1,q_2,q_3)$ is a three-dimensional
corner restricted by four planes and conditions: $q_1\geq0$, $q_2\geq0$, $q_3\geq0$,
and $q_1+q_2+q_3\leq1$.
Further, a representation of Eq.~(\ref{eq:rho_q1-4}) in Pauli matrices takes the
form
\begin{eqnarray}
   \label{eq:rho_q1-4_Bloch}
   \rho_{AB}&&=\frac{1}{4}\{1
   + (q_1+q_2+2q_3-1)(\sigma_z\otimes1
   + 1\otimes\sigma_z)\nonumber\\
   &&+ (q_1-q_2)(\sigma_x\otimes\sigma_x 
   + \sigma_y\otimes\sigma_y)
   + [1-2(q_1+q_2)]\sigma_z\otimes\sigma_z\}.
\end{eqnarray}
By the local unitary transformations, the state (\ref{eq:rho_q1-4}) is reduced to
\cite{PABJWK04,APVW07}
\begin{equation}
   \label{eq:rho_p1-4}
   \rho_{AB}=q^\prime_1|\Phi^+\rangle\langle\Phi^+| + q^\prime_2|\Phi^-\rangle\langle\Phi^-|
	 + q^\prime_3|01\rangle\langle01|
	 + q_4^\prime|10\rangle\langle10|,
\end{equation}
where $|\Phi^{\pm}\rangle=(|00\rangle\pm|11\rangle)/\sqrt{2}$ is the second pair of
Bell states.
The state (\ref{eq:rho_p1-4}) contains as a special case the Horodecki state
\cite{KHJP10} which is a mixture of one Bell state and separable states orthogonal to
the Bell one.
Last, the class of maximally discordant mixed states (MDMS) having maximal quantum
discord versus classical correlations is given by \cite{GGZ11}
\begin{equation}
   \label{eq:rho_mdms}
   \rho_{AB}=\epsilon|\Phi^+\rangle\langle\Phi^+|
	 + (1-\epsilon)(m|01\rangle\langle01|
	 + (1-m)|10\rangle\langle10|)
\end{equation}
($\epsilon$ and $m$ play a role of concentrations).

\section{Formulas for branches
}
\label{sect:Branches}
One-way quantum deficit for a bipartite state $\rho_{AB}$ is defined as the minimal
increase of entropy after a von Neumann measurement on one party (without loss of
generality, say, $B$)
\cite{HHHOSSS05} (see also, e.g., \cite{MBCPV12,BDSRSS18})
\begin{equation}
   \label{eq:D}
   {\rm\Delta}=\min_{\{\rm\Pi_k\}}S(\tilde\rho_{AB})-S(\rho_{AB}),
\end{equation}
where $S(\cdot)$ means the von Neumann entropy and
\begin{equation}
   \label{eq:rho_tilde}
   \tilde\rho_{AB}\equiv\sum_kp_k\rho^k_{AB}
	 =\sum_k(I\otimes{\rm\Pi}_k)\rho_{AB}(I\otimes{\rm\Pi}_k)^+
\end{equation}
is the weighted average of post-measured states
\begin{equation}
   \label{eq:rho-k}
   \rho^k_{AB}
	 =\frac{1}{p_k}(I\otimes{\rm\Pi}_k)\rho_{AB}(I\otimes{\rm\Pi}_k)^+
\end{equation}
with the probabilities
\begin{equation}
   \label{eq:p_k}
	 p_k={\rm Tr}(I\otimes{\rm\Pi}_k)\rho_{AB}(I\otimes{\rm\Pi}_k)^+.
\end{equation}
In Eq.~(\ref{eq:rho_tilde}), $\rm\Pi_k$ ($k=0,1$) are the general
orthogonal projectors
\begin{equation}
   \label{eq:Pi}
   {\rm\Pi}_k=V\pi_kV^+,
\end{equation}
where $\pi_k=|k\rangle\langle k|$ and transformations $\{V\}$ belong to
the special unitary group $SU(2)$.
Rotations $V$ may be parametrized by two angles $\theta$ and $\phi$
(polar and azimuthal, respectively):
\begin{equation}
   \label{eq:V}
   V
	 =\left(
      \begin{array}{cc}
      \cos(\theta/2)&-e^{-i\phi}\sin(\theta/2)\\
      e^{i\phi}\sin(\theta/2)&\cos(\theta/2)
      \end{array}
   \right)
\end{equation}
with $0\le\theta\le\pi$ and $0\le\phi<2\pi$.

Eigenvalues of matrix (\ref{eq:rhoXXZS}) are equal to
\begin{equation}
   \label{eq:lam1-4}
   \lambda_{1,2}=(1\pm2s_1+c_3)/4,\qquad \lambda_{3,4}=(1\pm2c_1-c_3)/4.
\end{equation}
Using these equations one gets the pre-measurement entropy function
\begin{equation}
   \label{eq:preS}
   S(s_1,c_1,c_3)
	 =h_4((1+2s_1+c_3)/4,(1-2s_1+c_3)/4,(1+2c_1-c_3)/4,(1-2c_1-c_3)/4),
\end{equation}
where
$h_4(x_1,x_2,x_3,x_4)=-x_1\log{x_1}-x_2\log{x_2}-x_3\log{x_3}-x_4\log{x_4}$
with condition $x_1+x_2+x_3+x_4=1$ is the quaternary entropy function.
In explicit form,
\begin{eqnarray}
   \label{eq:preSa}
   &&S(s_1,c_1,c_3)=2\ln2
	 \nonumber\\
	 &&-\frac{1}{4}[(1+2c_1-c_3)\ln(1+2c_1-c_3) + (1-2c_1-c_3)\ln(1-2c_1-c_3)
	 \nonumber\\
	 &&+ (1+2s_1+c_3)\ln(1+2s_1+c_3) + (1-2s_1+c_3)\ln(1-2s_1+c_3)].
\end{eqnarray}

The quantum state (\ref{eq:rhoXXZS}) after measurements ${\rm\Pi}_k$ ($k=0,1$) and
averaging in accordance with Eq.~(\ref{eq:rho_tilde}) takes the form
\begin{equation}
   \label{eq:tilde_rho}
   {\tilde\rho}_{AB}
  =\frac{1}{4}\!
	 \left(
      \begin{array}{cccc}
      1+s_1+(s_1&\bullet&\bullet&\bullet\\
      +c_3)\cos^2\!\theta&&&\\ \\
      \frac{1}{2}(s_1+c_3)e^{i\phi}\sin\!2\theta&1-c_3+(s_1&\bullet&\bullet\\
      &+c_3)\sin^2\!\theta&&\\ \\
      \frac{1}{2}c_1e^{i\phi}\sin\!2\theta&c_1\sin^2\!\theta&1-c_3-(s_1&\bullet\\
      &&-c_3)\sin^2\!\theta&\\ \\
      c_1e^{2i\phi}\sin^2\!\theta&\ -\frac{1}{2}c_1e^{i\phi}\sin\!2\theta&\
			\frac{1}{2}(s_1-c_3)e^{i\phi}\sin\!2\theta&\ 1-s_1-(s_1\\
			&&&-c_3)\cos^2\!\theta
      \end{array}
   \right)
\end{equation}
(for the sake of simplicity, the bold points are put instead of corresponding complex
conjugated matrix elements of the Hermitian matrix ${\tilde\rho}_{AB}$).

We used the Mathematica software to extract the eigenvalues of matrix
(\ref{eq:tilde_rho}).
First of all we established that the secular equation for the given matrix
is factorized into product of two polynomials of second orders.
In proving this key property, the following trick turned out useful.
Namely, we first transformed the matrix elements to the exponential form
(the command TrigToExp), found the secular equation, factorized it, then returned
by applying the command ExpToTrig to the trigonometric expressions and finally
simplified the result using the command FullSimplify.
As a result, we arrived at the eigenvalues of post-measurement state
(\ref{eq:tilde_rho}),
\begin{eqnarray}
   \label{eq:Lam}
	 \Lambda_{1,2}=\frac{1}{4}\lbrack\!\lbrack1+s_1\cos\theta\pm\{(s_1+c_3\cos\theta)^2
	 +c_1^2\sin^2\!\theta\}^{1/2}\rbrack\!\rbrack,
	 \nonumber\\ \\
	 \Lambda_{3,4}=\frac{1}{4}\lbrack\!\lbrack1-s_1\cos\theta\pm\{(s_1-c_3\cos\theta)^2
	 +c_1^2\sin^2\!\theta\}^{1/2}\rbrack\!\rbrack.
	 \nonumber
\end{eqnarray}
It is seen that the azimuthal angle $\phi$ has dropped out from the given expressions.
This is a general property and occurs every time 
when only one of the off-diagonals of the density matrix is non-zero \cite{VR12,MHR15}.
Hence, the optimization reduces to that in a single variable $\theta$.

Using Eqs.~(\ref{eq:Lam}) we find the post-measured entropy (entropy after measurement)
\begin{equation}
   \label{eq:postS}
   \tilde S(\theta)\equiv S(\tilde\rho_{AB})=h_4(\Lambda_1,\Lambda_2,\Lambda_3,\Lambda_4),
\end{equation}
where $h_4(\cdot)$ is again the quaternary entropy function.
The function $\tilde S$ of argument $\theta$ is invariant under
the transformation $\theta\to\pi-\theta$ therefore it is enough to restrict oneself by
values of $\theta\in[0,\pi/2]$.
Moreover, the pre- and post-measured entropies $S$ and $\tilde S$, as functions
of $s_1$ and $c_1$, are symmetric under the reflections
$s_1\to -s_1$ and $c_1\to -c_1$.

Notice the following.
The post-measument entropy $\tilde S$ is related to the conditional entropy
$S_{cond}$ [see Eqs.~(25)-(26) in Ref.~\cite{Y17}] by equation (see \cite{SXL13} and
also
Appendix in Ref.~\cite{Y17})
\begin{equation}
   \label{eq:postS-Scond}
   \tilde S(\theta)=S_{cond}(\theta)+h((1+s_1\cos\theta)/2),
\end{equation}
with $h(x)=-x\ln x-(1-x)\ln(1-x)$ being Shennon's binary entropy function.
The first derivative of $h$-term with respect to $\theta$ equals zero
at both endpoints $\theta=0$ and $\pi/2$.
Hence if the first derivative of $S_{cond}(\theta)$ with respect to $\theta$ vanishes
at $\theta=0$ and $\pi/2$ then the same takes place for the post-measured entropy
$\tilde S(\theta)$ and vice versa.
Moreover, in accordance with Eq.~(\ref{eq:postS-Scond}), the non-optimized one-way
deficit can be written as
\begin{equation}
   \label{eq:D-Scond}
   \Delta(\theta)=h((1+s_1\cos\theta)/2)-S(\rho_{AB})+S_{cond}(\theta).
\end{equation}
On the other hand, the measurement-dependent quantum discord can be presented as
\begin{equation}
   \label{eq:Q-Scond}
   Q(\theta)=h((1+s_1\cos\theta)/2)|_{\theta=0}-S(\rho_{AB})+S_{cond}(\theta)
\end{equation}
because $h((1+s_1\cos\theta)/2)|_{\theta=0}=S(\rho_B)$, where
$\rho_B={\rm Tr}_A\rho_{AB}$.
So, Eqs.~(\ref{eq:D-Scond}) and (\ref{eq:Q-Scond}) show a very close mathematical
relation between both measures of quantum correlation: the only difference is either
we take $h$-term with arbitrary $\theta$ or set $\theta=0$.
This relation is valid for general X states.

From Eqs.~(\ref{eq:Lam}) and (\ref{eq:postS}), one gets an expression
for the post-measurement entropy:
\begin{eqnarray}
   \label{eq:tildeSs1c1c3}
   &&\tilde S(\theta; s_1,c_1,c_3)=2\ln2
	 \nonumber\\
	 &&-\frac{1}{4}\sum_{m,n=1}^2\biggl(1+(-1)^m s_1\cos\theta
	 +(-1)^n\sqrt{(s_1+(-1)^m c_3\cos\theta)^2+c_1^2\sin^2\theta}\biggr)
	 \nonumber\\
	 &&\times\ln\biggl(1+(-1)^m s_1\cos\theta+(-1)^n\sqrt{(s_1
	 +(-1)^m c_3\cos\theta)^2+c_1^2\sin^2\theta}\biggr).
\end{eqnarray}
In the following analysis, this function will serve us to probe and identify the types
of subregions in phase diagrams.
The function $\tilde S(\theta; s_1,c_1,c_3)$ is differentiable at any point $\theta$
and, in full conferment with the statement made in the Introduction,
its  first derivatives with respect to $\theta$ identically
equal zero for $\forall\,s_1,c_1,c_3\in{\cal T}$
at both ends of the interval $[0,\pi/2]$.

The post-measurement entropy at the endpoint $\theta=0$ is given as
\begin{eqnarray}
   \label{eq:S0s1c1c3}
   &&\tilde S_0\equiv\tilde S(0)=2\ln2-\frac{1}{2}(1-c_3)\ln(1-c_3)
	 \nonumber\\
	 &&-\frac{1}{4}[(1+2s_1+c_3)\ln(1+2s_1+c_3) + (1-2s_1+c_3)\ln(1-2s_1+c_3)]
\end{eqnarray}
and at the endpoint $\theta=\pi/2$:
\begin{eqnarray}
   \label{eq:S1s1c1c3}
   &&\tilde S_{\pi/2}\equiv\tilde S(\pi/2)=2\ln2
	 -\frac{1}{2}\biggl[\biggl(1+\sqrt{s_1^2+c_1^2}\biggr)\ln\biggl(1+\sqrt{s_1^2+c_1^2}\biggr)
	 \nonumber\\
	 &&+\biggl(1-\sqrt{s_1^2+c_1^2}\biggr)\ln\biggl(1-\sqrt{s_1^2+c_1^2}\biggr)\biggr].
\end{eqnarray}

Equations~(\ref{eq:preS}),(\ref{eq:Lam}), and (\ref{eq:postS}) define the
measurement-dependent one-way deficit function
\begin{equation}
   \label{eq:D_theta}
   \Delta(\theta)={\tilde S}(\theta)-S.
\end{equation}
In the wake of ${\tilde S}(\theta)$, the first derivative of $\Delta(\theta)$
with respect to $\theta$ is identically equal to zero at both endpoints $\theta=0$ and
$\theta=\pi/2$:
\begin{equation}
   \label{eq:postSD1}
   {\tilde S}^{\prime}(0)=\Delta^{\prime}(0)\equiv0,\qquad
   {\tilde S}^{\prime}(\pi/2)=\Delta^{\prime}(\pi/2)\equiv0.
\end{equation}

By analogy with the quantum discord \cite{Y14,Y14a,MHR15,Y15}
(see also \cite{Y17}), the one-way
quantum deficit may be written as almost closed analytical formula \cite{Y18}
\begin{equation}
   \label{eq:rmD}
   \rm\Delta=\min\{\Delta_0,\Delta_{\pi/2},\Delta_\vartheta\}.
\end{equation}
Thus, it can consist of three branches.
However, it should be noted that in general these branches may further out split into
new subbranches.

Using the above equations one obtains the expression for the 0-branch:
\begin{eqnarray}
   \label{eq:D0s1c1c3}
   &&\Delta_0(s_1,c_1,c_3)=-\frac{1}{2}(1-c_3)\ln(1-c_3)
	 \nonumber\\
	 &&+\frac{1}{4}[(1+2c_1-c_3)\ln(1+2c_1-c_3) + (1-2c_1-c_3)\ln(1-2c_1-c_3)].
\end{eqnarray}
Surprisingly, this branch is identical to the similar branch of discord
$Q_0$ \cite{Y17}, i.e., $\Delta_0=Q_0$.
Moreover the function $\Delta_0$ is symmetric under the reflection $c_1\to-c_1$
and does not depend on $s_1$.

For the branch $\Delta_{\pi/2}$ with the $\pi/2$ optimal measured angle,
we have
\begin{eqnarray}
   \label{eq:Dpi/2s1c1c3}
   &&\Delta_{\pi/2}(s_1,c_1,c_3)
	 =-\frac{1}{2}\biggl[\biggl(1+\sqrt{s_1^2+c_1^2}\biggr)\ln\biggl(1+\sqrt{s_1^2+c_1^2}\biggr)
	 \nonumber\\
	 &&+\biggl(1-\sqrt{s_1^2+c_1^2}\biggr)\ln\biggl(1-\sqrt{s_1^2+c_1^2}\biggr)\biggr]
	 \nonumber\\
	 &&+\frac{1}{4}[(1+2c_1-c_3)\ln(1+2c_1-c_3) + (1-2c_1-c_3)\ln(1-2c_1-c_3)
	 \nonumber\\
	 &&+ (1+2s_1+c_3)\ln(1+2s_1+c_3) + (1-2s_1+c_3)\ln(1-2s_1+c_3)].
\end{eqnarray}
This function is symmetric both on $s_1$ and $c_1$.

The third and last branch $\Delta_\vartheta$ is obtained by numerically solving
the one-dimensional optimization problem for the function $\tilde S(\theta)$.

We turn now to the discussion of boundaries distinguishing the phases of quantum
correlation.

\section{Equations for boundaries}
\label{sect:Bonds}
The problem of phases and boundaries between them is well known in thermodynamics and
statistical physics \cite{B82,LL_StPh} as well as in other fields of science.
Classification of phases depends on characteristic features which are put in its
ground.
Therefore it is not surprising that, e.g., one discovers more and more new phases
for water.

Quantum correlations are piecewise-defined functions.
A choice of a certain branch in the given point of parameter space is dictated by
a minimum condition like (\ref{eq:rmD}).
But how to decompose the whole domain of definition of quantum correlation function
into subdomains?
The answer to this question was firstly given in Refs.~\cite{Y14,Y14a}
(see also \cite{MHR15,Y15}) with the quantum discord.

Applying those ideas to the one-way quantum deficit, one may say the following.
On the one hand, the boundary between the phases with zero and $\pi/2$ optimal
measurement angles is controlled by the equilibrium  condition
\begin{equation}
   \label{eq:S_0S_pi/2}
   \tilde S_0=\tilde S_{\pi/2}\quad
   {\rm or}\quad
   \Delta_0=\Delta_{\pi/2}.
\end{equation}
When crossing this boundary, the optimal measurement angle experiences
the jump $\Delta\vartheta=\pi/2$.

On the other hand, the transitions from the $\vartheta$-phase to the 0- or
$\pi/2$-one can occur via a bifurcation of minimum of the post-measurement
entropy function at the endpoints $\theta=0$ and $\pi/2$.
In these cases the equations for the boundaries are reduced to a requirement of
vanishing the second derivatives of post-measurement entropy or measurement-dependent
one-way deficit with respect to $\theta$ at corresponding endpoints \cite{YWF16,Y18}
\begin{equation}
   \label{eq:D0_11}
   \tilde S^{\prime\prime}(0)=0\quad
   {\rm or}\quad
   \Delta^{\prime\prime}(0)=0\quad
\end{equation}
for the 0-boundary and
\begin{equation}
   \label{eq:D1_11}
	 \tilde S^{\prime\prime}(\pi/2)=0\quad
   {\rm or}\quad
   \Delta^{\prime\prime}(\pi/2)=0
\end{equation}
for the $\pi/2$-one.
Here the optimal measurement angle jumps equal $\Delta\vartheta=0$.

For the quantum states under discussion the second derivatives at endpoints
follow from Eq.~(\ref{eq:tildeSs1c1c3}) and equal
\begin{eqnarray}
   \label{eq:S110}
   \tilde S^{\prime\prime}(0)=&&\frac{1}{4}\biggl\{s_1\ln\frac{1+2s_1+c_3}{1-2s_1+c_3}
	 +c_3\ln\frac{(1+2s_1+c_3)(1-2s_1+c_3)}{(1-c_3)^2}
	 \nonumber\\
	 &&-c_1^2\biggl[\frac{1}{s_1+c_3}\ln\frac{1+2s_1+c_3}{1-c_3}
	 +\frac{1}{s_1-c_3}\ln\frac{1-c_3}{1-2s_1+c_3}\biggr]\biggl\}
\end{eqnarray}
and
\begin{eqnarray}
   \label{eq:S11P}
	 \tilde S^{\prime\prime}(\pi/2)=c_1^2\frac{s_1^2+c_1^2-c_3^2}{2r^3}\ln\frac{1+r}{1-r}
	 -\frac{1}{2}s_1^2\biggl(\frac{(1+c_3/r)^2}{1+r}+\frac{(1-c_3/r)^2}{1-r}\biggr)
\end{eqnarray}
with $r=\sqrt{s_1^2+c_1^2}$.

Equations~(\ref{eq:S_0S_pi/2}), (\ref{eq:D0_11}), and (\ref{eq:D1_11})
together with expressions (\ref{eq:S0s1c1c3}), (\ref{eq:S1s1c1c3}),
(\ref{eq:D0s1c1c3}), (\ref{eq:Dpi/2s1c1c3}), (\ref{eq:S110}), and (\ref{eq:S11P})
are transcendental and they can be solved numerically by the bisection method.

The above mechanism of arising the boundaries between phases is confirmed for the
cases when the post-measurement entropy function is unimodal.
However, most recently \cite{Y18} a bimodal behavior for the post-measurement entropy
function has been found and a new important equation has been added to the collection
of boundary equations, namely
\begin{equation}
   \label{eq:0prime}
   \tilde S_0=\tilde S_{\vartheta}\quad
   {\rm or}\quad
   \Delta_0=\Delta_{\vartheta}.
\end{equation}
These conditions reflect a jump (sudden change) of optimal measurement angle from
zero to a finite step $\Delta\vartheta>0$ but $\Delta\vartheta\neq\pi/2$.
For definiteness, we will call such a boundary as the 0$^\prime$-one.

Generally speaking the jumps between the endpoint $\theta=\pi/2$ and global interior
minimum are possible theoretically but up to the present they were not met in our
practice.

Note for a reference that we solved the equations (\ref{eq:0prime}) for the
0$^\prime$-boundary also by the bisection method with searching the interior minimum
of post-measurement entropy or measurement-dependent one-way quantum deficit
at every step of iteration procedure by the golden section method.

The above classification of phases is based on the type of optimal measurement angle.
But in general such an approach does not exhaust all possible branches of
piecewise-defined function.
New subbranches can appear from the splitting of some original branches.
For instance, such a phenomenon was previously observed for the $\pi/2$-branch of
discord, $Q_{\pi/2}$, which decays into two subbranches $Q_{\pi/2}^{(1)}$ and
$Q_{\pi/2}^{(2)}$ in the limit of Bell-diagonal states
(see Fig.~7b in Ref.~\cite{Y15} and Sec.~2.1 in Ref.~\cite{Y17}).
The reason is simple: when $s_1=s_2=0$, the function $Q_{\pi/2}(c_1,c_2,c_3)$ by
$|c_1|\neq|c_2|$ contains inside itself the piecewise functions like $|c_1-c_2|$.
In similar situations, the additional boundaries are revealed from the detection
of fracture points, i.e., from the condition of discontinuity for the first
derivatives with respect to the parameters of the model:
\begin{equation}
   \label{eq:neqDprime}
   {\rm\Delta}_x^\prime(x-0)\neq{\rm\Delta}_x^\prime(x+0),\quad x\in\{s_1,c_1,c_3\}.
\end{equation}
However, we did not observe such cases in the present paper because $c_1=c_2$.

\section{Phase diagram for the symmetric XXZ state}
\label{sect:PhDiagr}
The task now is to separate the body $\cal T$ by using boundary equations
into subdomains each one corresponding to a certain branch according to the
minimum condition (\ref{eq:rmD}).
Our strategy is reduced to the following.
We will take different two-dimensional sections of tetrahedron $\cal T$, solve the
equations for boundaries, and then visually analyze the shapes of curves
$\tilde S(\theta)$ or $\Delta(\theta)$.
The position of global minimum of these curves will give us the answer to the question
about the type of branch in the given part of section.

The one-way deficit function ${\rm\Delta}(s_1,c_1,c_3)$ is invariant under the
transformations $s_1\to-s_1$ and $c_1\to-c_1$.
Hence the body $\cal T$ with all subdomains is symmetric relative to
the mirror reflections in the planes $s_1=0$ and $c_1=0$.
Owing to this, it is enough to study the phases only in a quarter of tetrahedron
$\cal T$.

\subsection{Phase diagrams on faces}
\label{sect:PhDiagrFace}
\begin{figure}[t]
\begin{center}
\epsfig{file=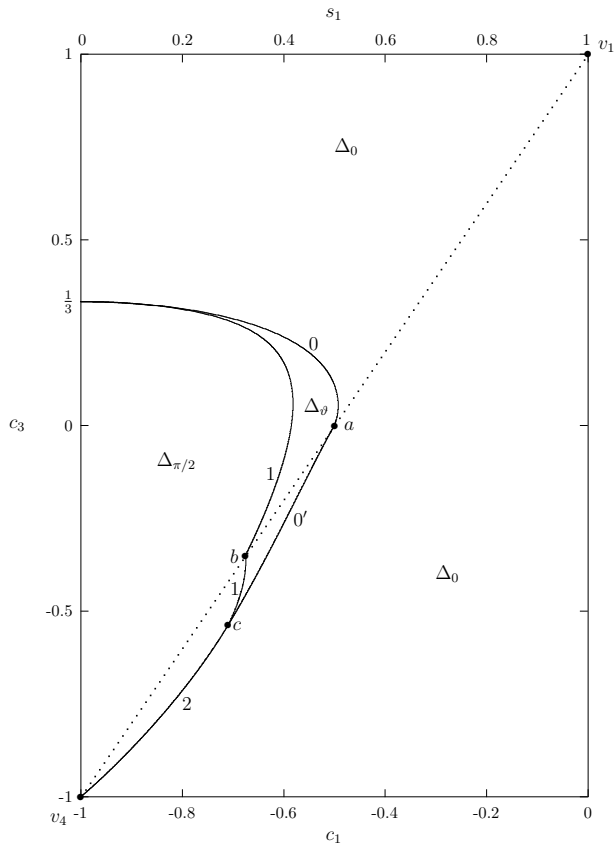,width=8cm}
\caption{
Phase diagram of one-way quantum deficit on the faces of tetrahedron $\cal T$.
The diagonal dotted line $v_1v_4$ dividing the figure area in two triangle parts
is the corresponding edge of the tetrahedron.
The upper triangle is a half of the face $v_1v_2v_4$ while
the lower one is a half of the face $v_1v_3v_4$.
(The latter represents the phase diagram \cite{Y18} in new variables.)
The points ``a'' and ``b'' lie on the diagonal $v_1v_4$, their coordinates
$(c_1,c_3)$ being equal $(-0.5,0)$ and $(-0.675151,-0.350302)$, respectively,
whereas the point ``c'' is located at $(-0.709723,-0.538191)$
}
\label{fig:zzf}
\end{center}
\end{figure}

We begin the analysis with the faces of tetrahedron $\cal T$.
Due to the mirror symmetry, it is enough to study the phases only on two adjacent
semi-faces.
Without loss of generality, we consider the faces $v_1v_2v_4$ and $v_1v_3v_4$
(see Fig.~\ref{fig:z_xxz-m2}).
Both halves of these two adjacent faces are shown in Fig.~\ref{fig:zzf}
in an unfolded form.

\begin{figure}[t]
\begin{center}
\epsfig{file=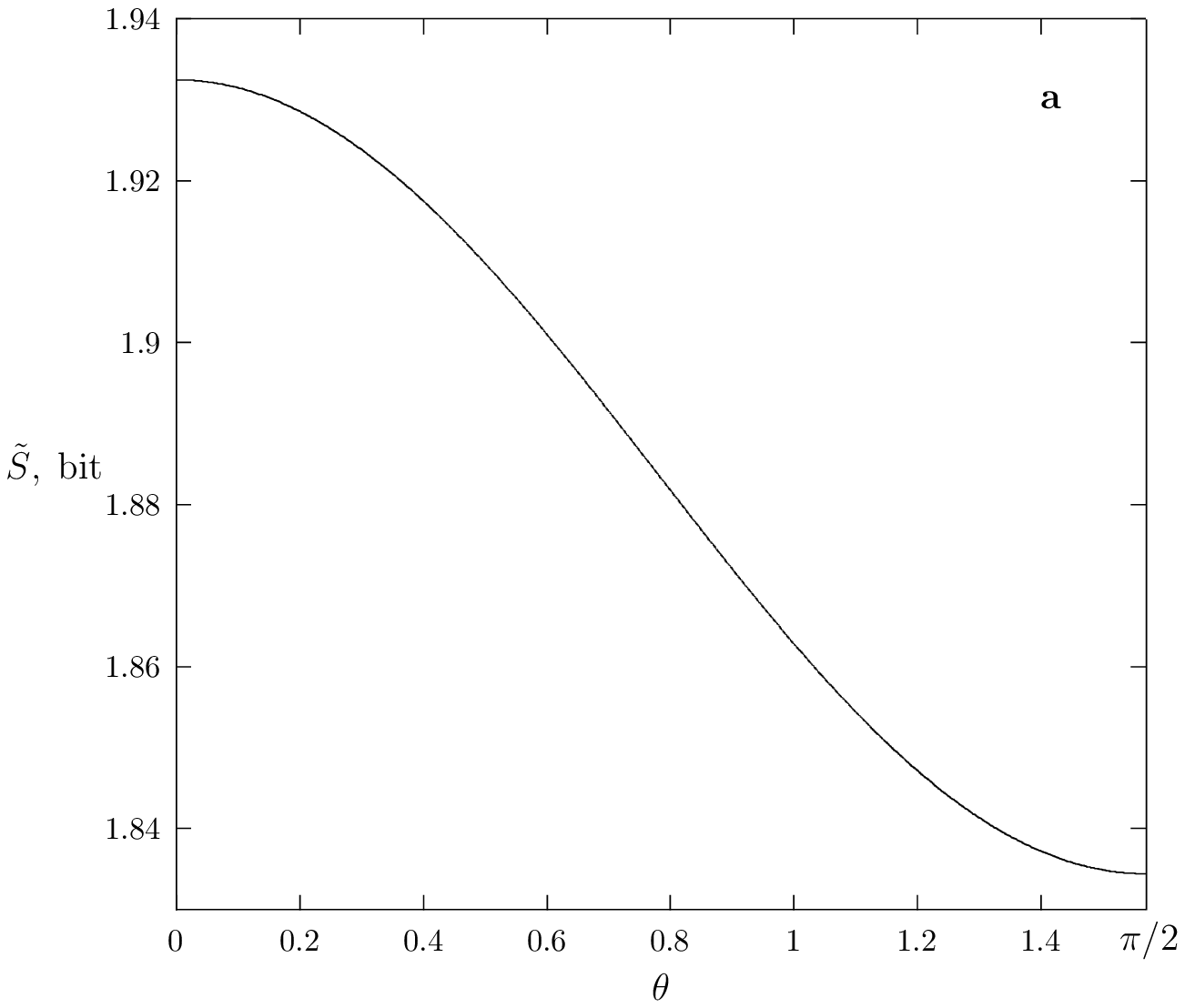,width=5.5cm}
\vspace{0.5cm}
\hspace{0.5cm}
\epsfig{file=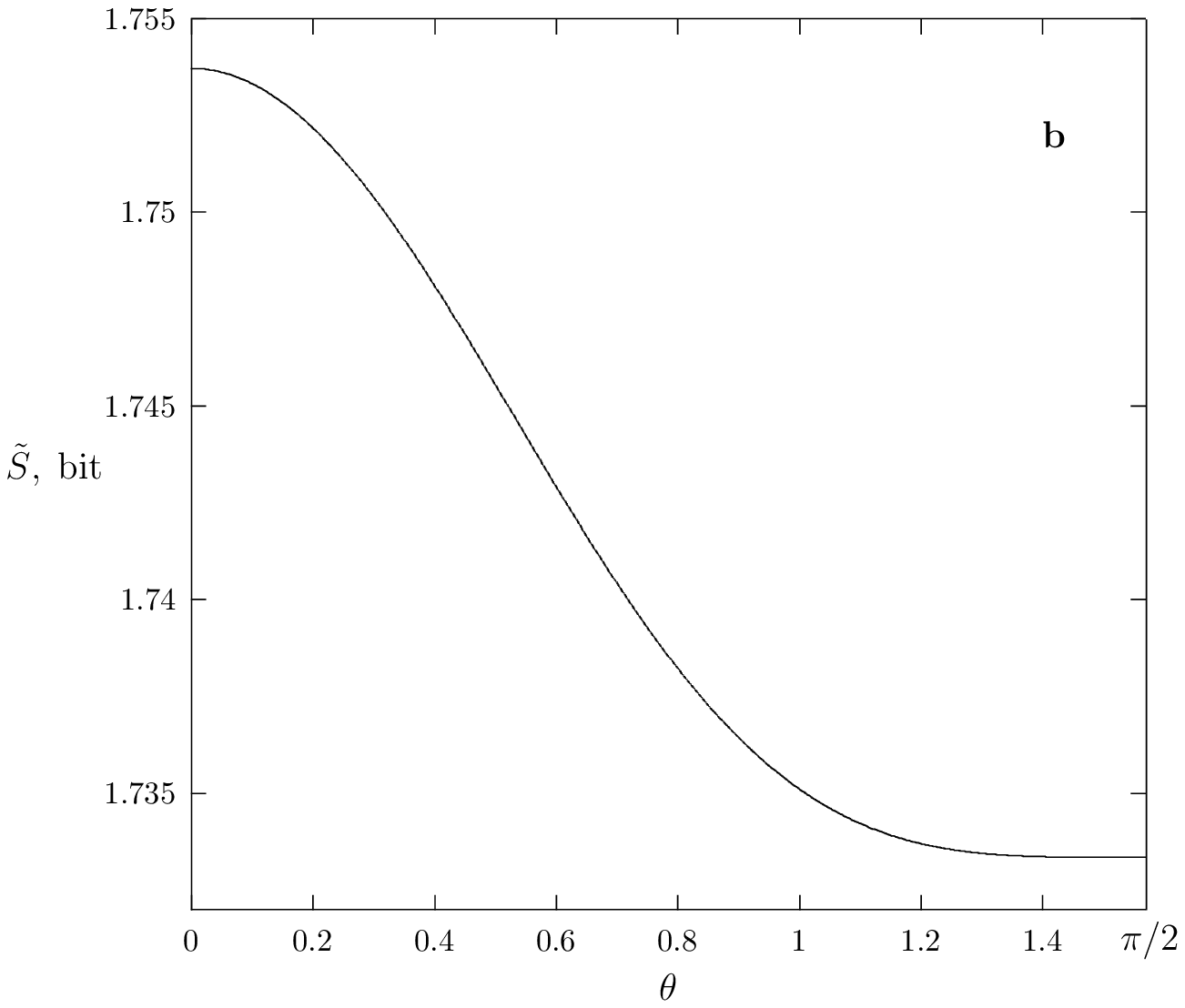,width=5.5cm}
\epsfig{file=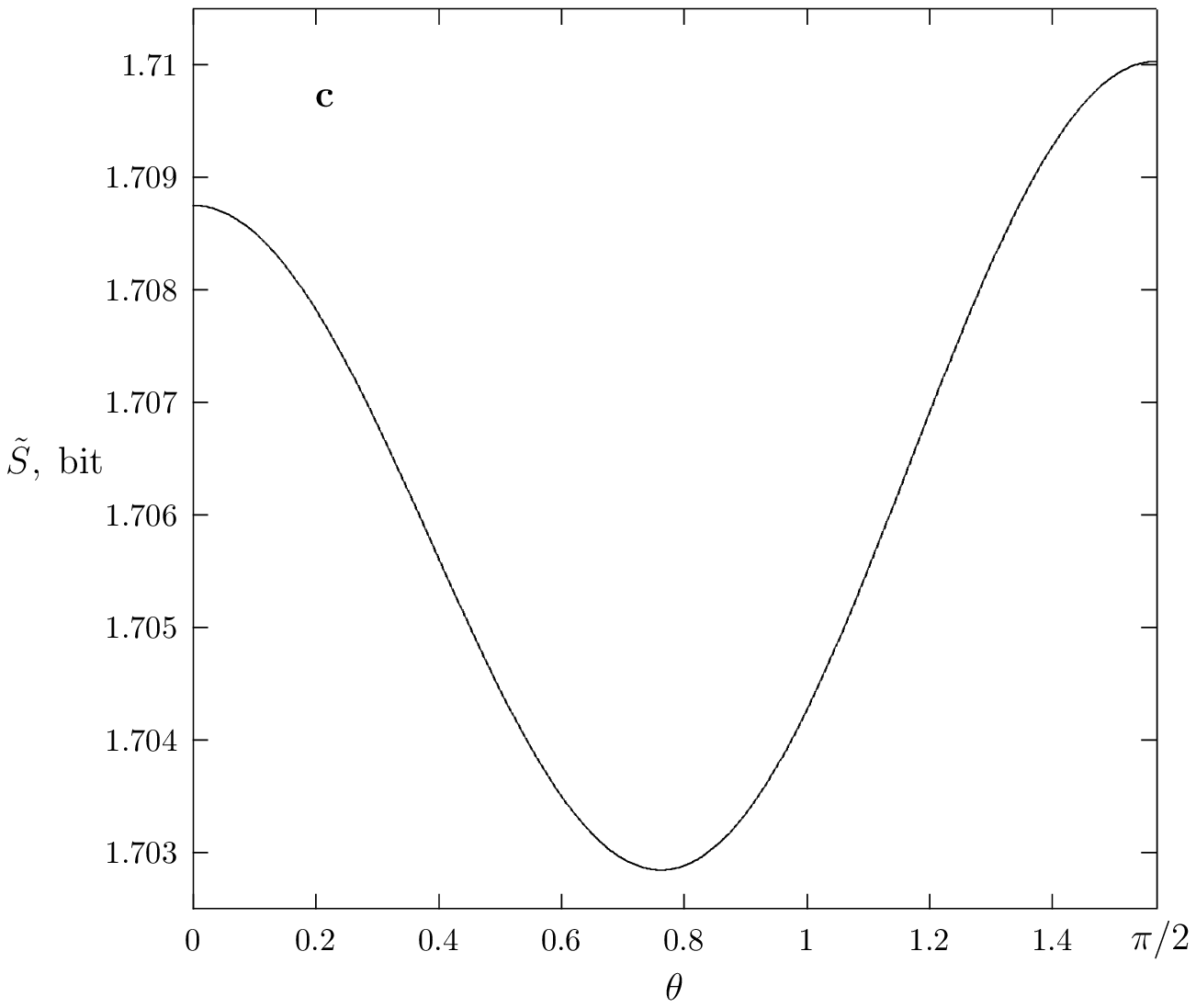,width=5.5cm}
\hspace{0.5cm}
\epsfig{file=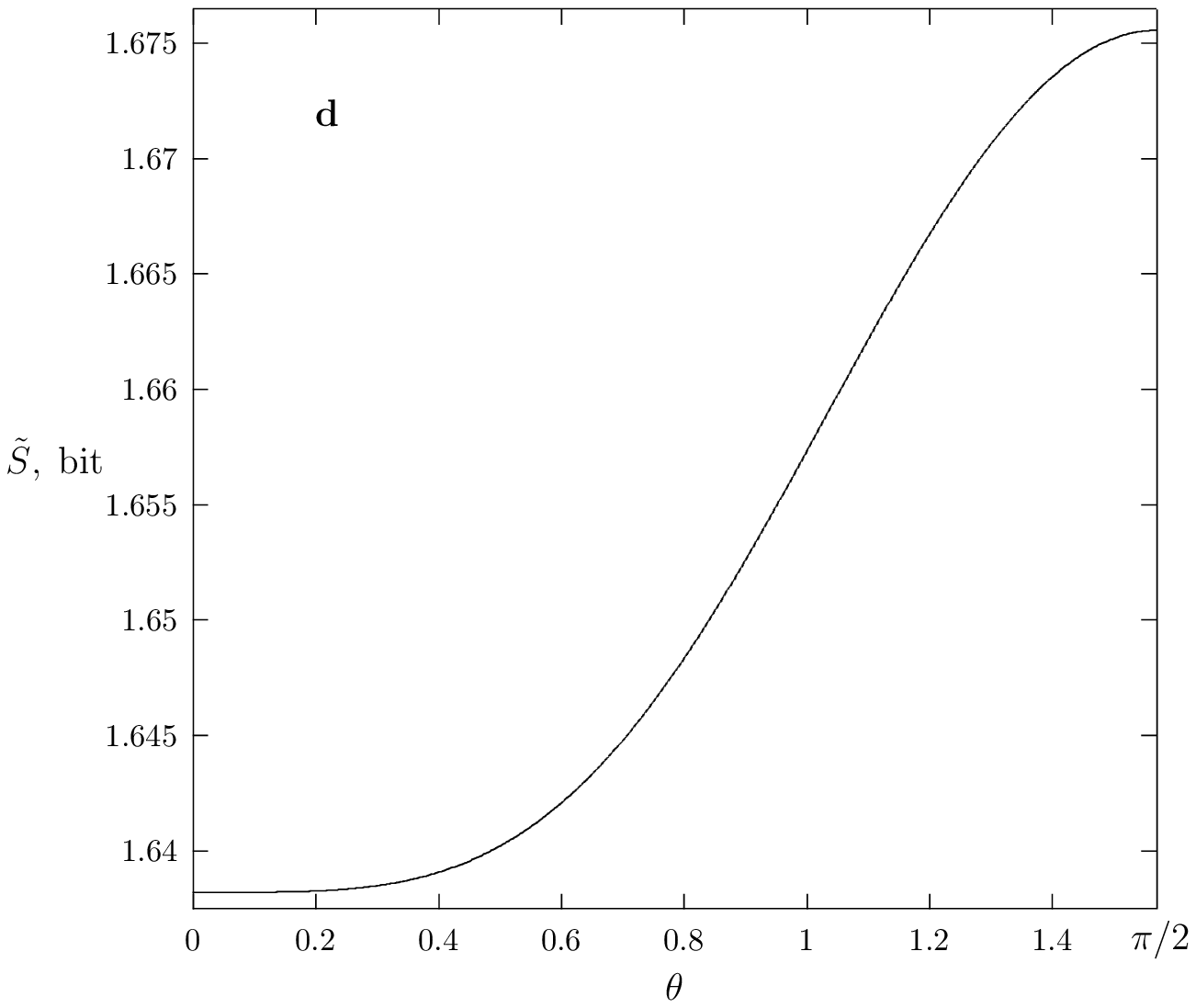,width=5.5cm}
\caption{
Evolution of post-measurement entropy shapes by pass along the line $c_3=0.15$
on the face $v_1v_2v_4$ [i.e, when $c_1=(1-c_3)/2$)] for $s_1=0.2$~(a), 0.406975~(b),
0.44~(c), and 0.483997~(c).
Cases (b) and (d) correspond to the moments of bifurcations of minimum which occur
on the boundary curves 1 and 0, respectively
}
\label{fig:zs015}
\end{center}
\end{figure}

Consider first the upper triangle.
Here $c_1=-(1-c_3)/2$.
Solution of equations from the previous section leads to the curves labeled in
Fig.~\ref{fig:zzf} by symbols 0 and 1 which correspond to the 0- and
$\pi/2$-boundaries, respectively.
Testing the different parts of triangle by the shape of curve $\tilde S(\theta)$
allows to identify the types of separate subdomains;
they are marked in Fig.~\ref{fig:zzf} by corresponding names of branches.
Three found types of post-measurement entropy shapes
(monotonically decreasing, unimodal, and monotonically increasing)
are illustrated in Fig.~\ref{fig:zs015}.
More complicated shapes of $\tilde S(\theta)$ are absent on this face.

Phase diagram on the lower triangle part
of Fig.~\ref{fig:zzf} [here $s_1=(1+c_3)/2$] corresponds to the case
which has been considered in Ref.~\cite{Y18}~(Fig.~7 there);
the only difference is the coordinates $(c_1,c_3)$ instead  $(q_1,q_2)$.
Here the bimodal behavior of $\tilde S(\theta)$ takes place and,
correspondingly, the $0^\prime$-boundary on which the optimal measurement
angle discontinuously changes its value exists between the points ``a'' and ``c''.
The curve $0^\prime$ is a line of continuously varying $\Delta\vartheta$ in the limits
from 0 to $\pi/2$ (see Table~1 in \cite{Y18}).
The critical line 2 corresponds to the boundary when $\Delta_0=\Delta_{\pi/2}$.

\begin{figure}[t]
\begin{center}
\epsfig{file=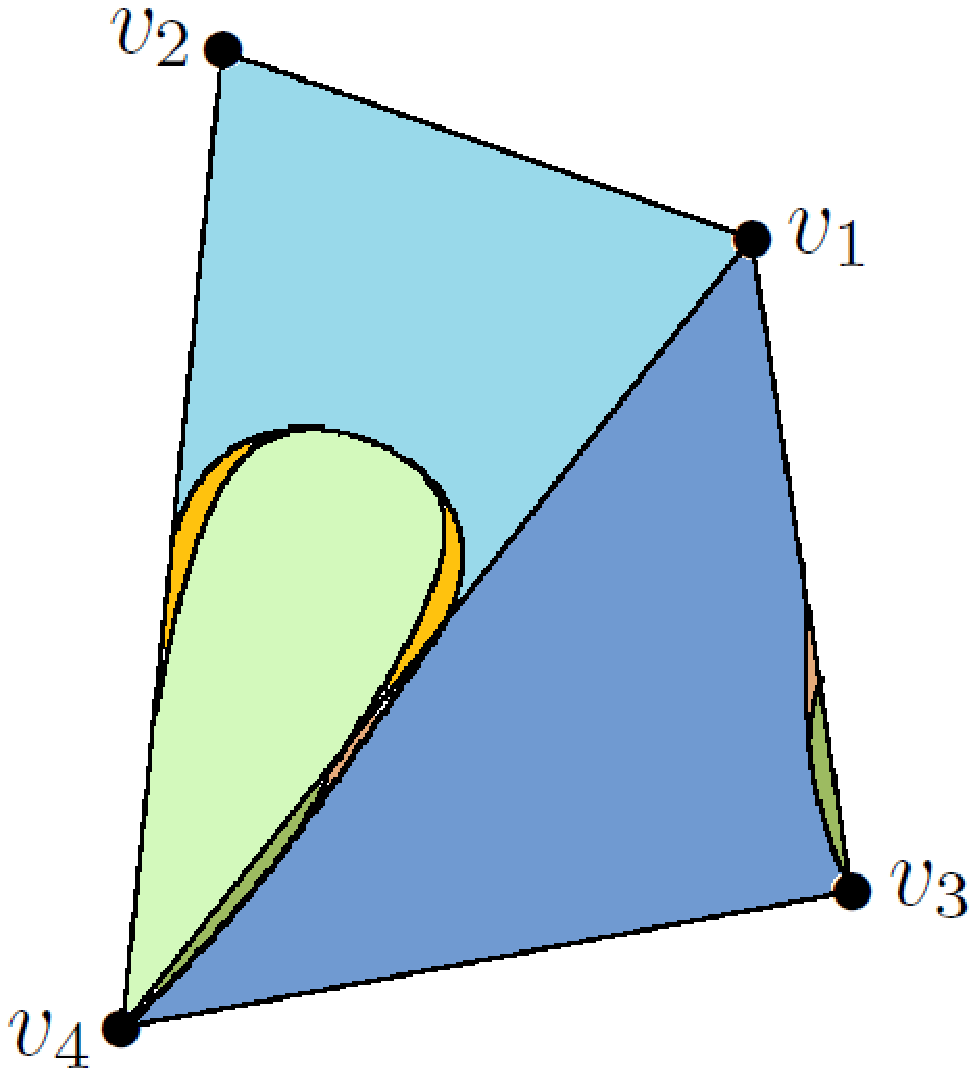,width=5.1cm}
\hspace{1.5cm}
\epsfig{file=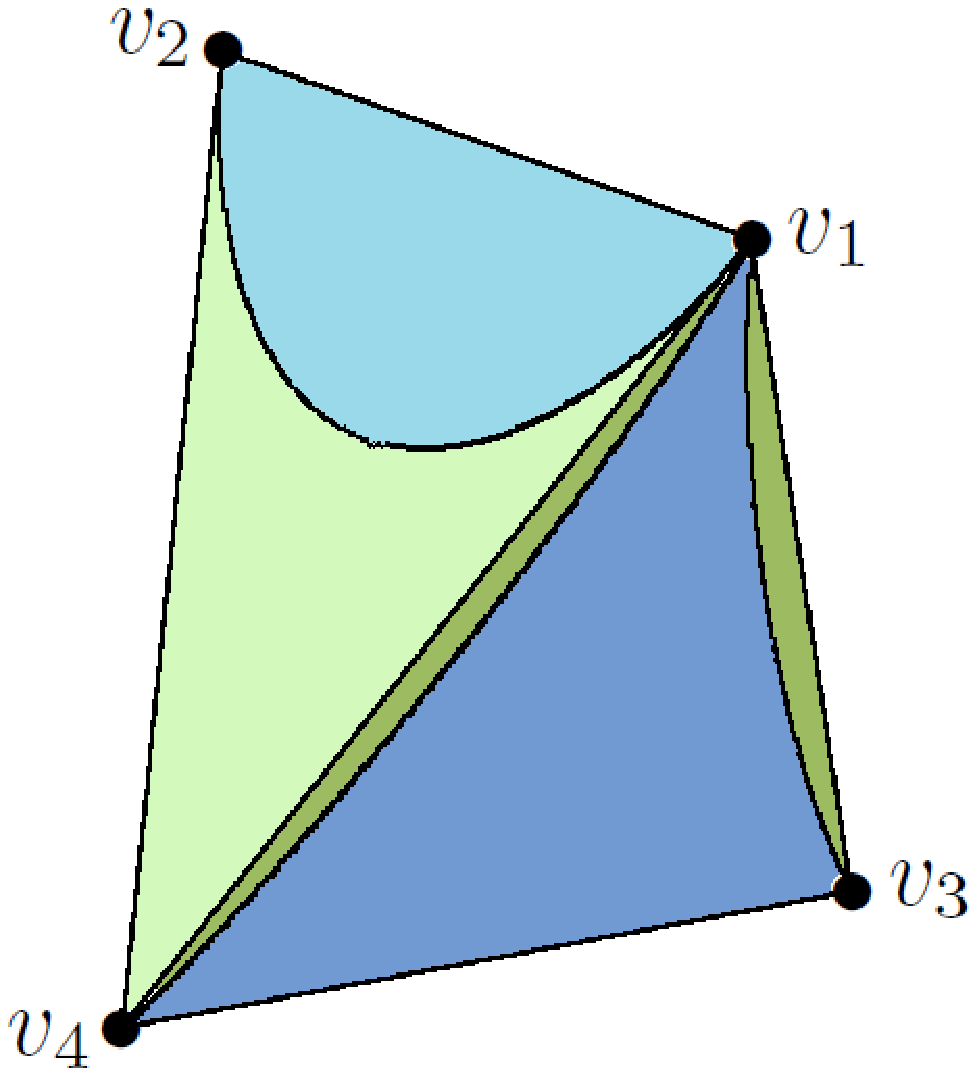,width=5.1cm}
\caption{
(Color online) An outward appearance of tetrahedron $\cal T$ for the one-way deficit
(left, see also Fig.~\ref{fig:zzf}) and discord (right).
The regions $\Delta_0$ and $Q_0$ are shown by the blue color while the $\Delta_{\pi/2}$
and $Q_{\pi/2}$ ones are shown by the green color.
The region $\Delta_{\vartheta}$ is yellow-colored and because
the region $Q_{\theta^*}$ lies inside the tetrahedron it is not seen
}
\label{fig:z_xxzb2a}
\end{center}
\end{figure}

It is interesting to compare the location of phases $\Delta_0$, $\Delta_{\pi/2}$,
and $\Delta_{\vartheta}$ in the tetrahedron $\cal T$ with that of similar
discord phases $Q_0$, $Q_{\pi/2}$, and $Q_{\theta^*}$ found earlier for the same
quantum states \cite{Y17}.
One can see the phases of one-way deficit and discord on faces of $\cal T$
in Fig.~\ref{fig:z_xxzb2a}.
Because the one-way quantum deficit must be identical to the quantum discord for the
Bell-diagonal states,
we satisfy ourselves that really ${\rm\Delta}=Q$ in the case $s_1=0$.
More and above, we observe that, due to the equality $\Delta_0=Q_0$, the one-way
deficit $\rm\Delta$ and discord $Q$ are equal when the  $\Delta_0$ and $Q_0$ regions
coincide.
This circumstance considerably enlarges the common part of total domain $\cal T$,
where both measures of quantum correlation yield the same result.

\subsection{Phase diagrams inside the tetrahedron $\cal T$}
\label{sect:PhDiagrCross}
Consider now the phase structure in the interior of tetrahedron.
The study will be performed by scanning the body $\cal T$ by taking cross-sections
with planes $c_3=const$.
We will go from the top to the bottom of the tetrahedron.

When $c_3=1$, i.e., on the edge $v_1v_2$, all quantum correlations vanish because
here $c_1=0$ and the system is purely classical.

In the subsequent investigation it will be convenient to consider five separate
intervals for $c_3$ taking into account the variation of phases in longitudinal
direction (see Fig.~\ref{fig:zzf}).

First of all the calculations show that the one-way deficit equals
${\rm\Delta}=\Delta_0$ in the band $1/3\leq c_3\leq1$.

\begin{figure}[t]
\begin{center}
\epsfig{file=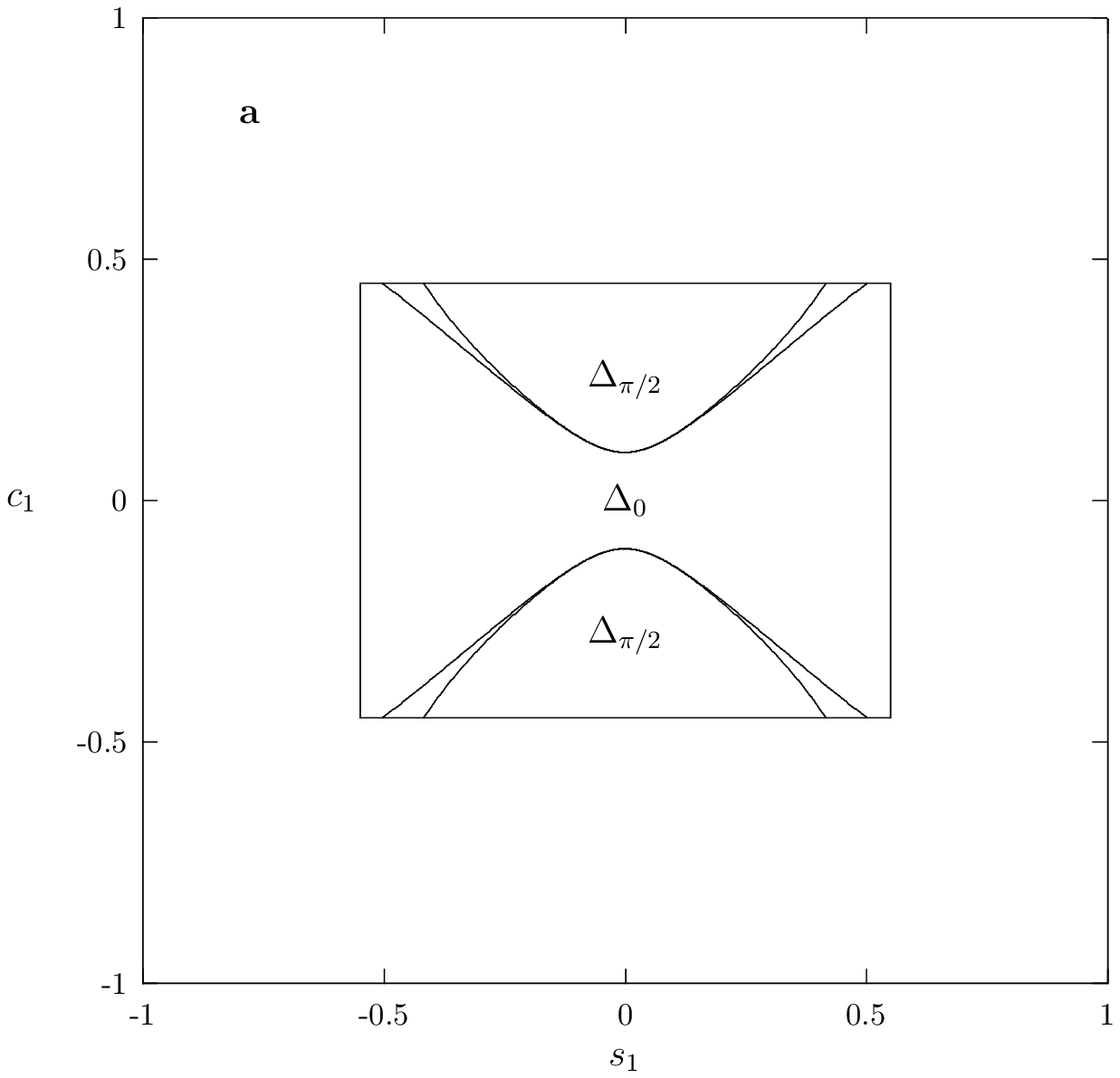,width=5.5cm}
\hspace{0.5cm}
\epsfig{file=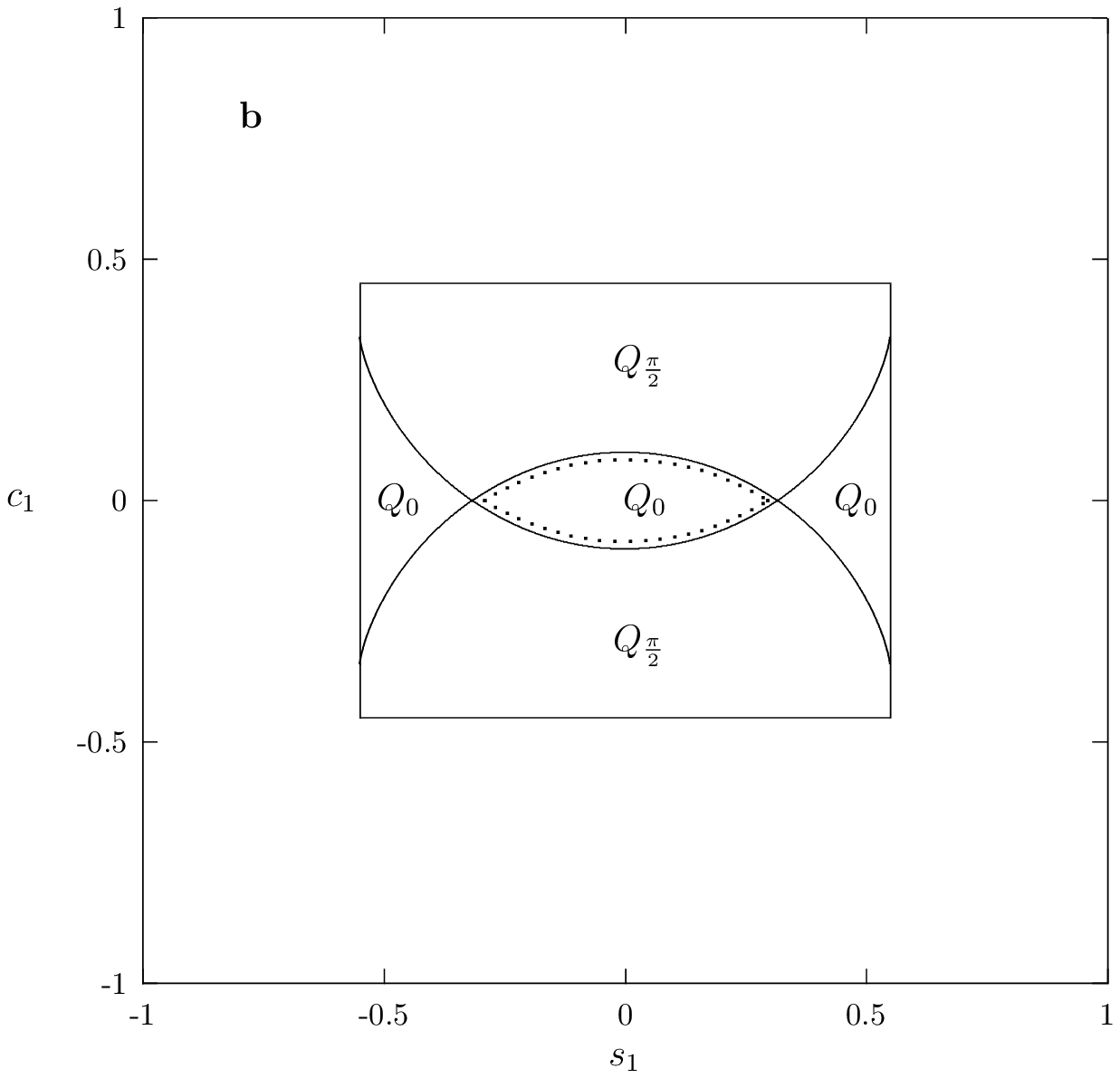,width=5.5cm}
\caption{
Phase diagrams of one-way deficit~(a) and discord~(b) by $c_3=0.1$.
Subregions $\Delta_{\vartheta}$ lie between the pairs of solid lines in a graph~(a).
Similar subregions for the quantum discord, $Q_{\theta^*}$, are thin too
and therefore their locations are shown only schematically by double
solid-dotted lines in a graph~(b) (see also Fig.~3 in \cite{Y17})
}
\label{fig:pd01}
\end{center}
\end{figure}

By $c_3<1/3$, the other two phases appear in the slices.
Typical phase diagram for the one-way deficit in the interval of $c_3$ from 1/3 to 0
is shown in Fig.~\ref{fig:pd01}a for the cross-section $c_3=0.1$.
Phase diagram for the discord is shown in Fig.~\ref{fig:pd01}b for a comparison.
One sees the significant differences between both measures of quantum correlation.
However they are the same on the line $s_1=0$ (Bell's case) and in common parts
of regions $\Delta_0$ and $Q_0$.

\begin{figure}[t]
\begin{center}
\epsfig{file=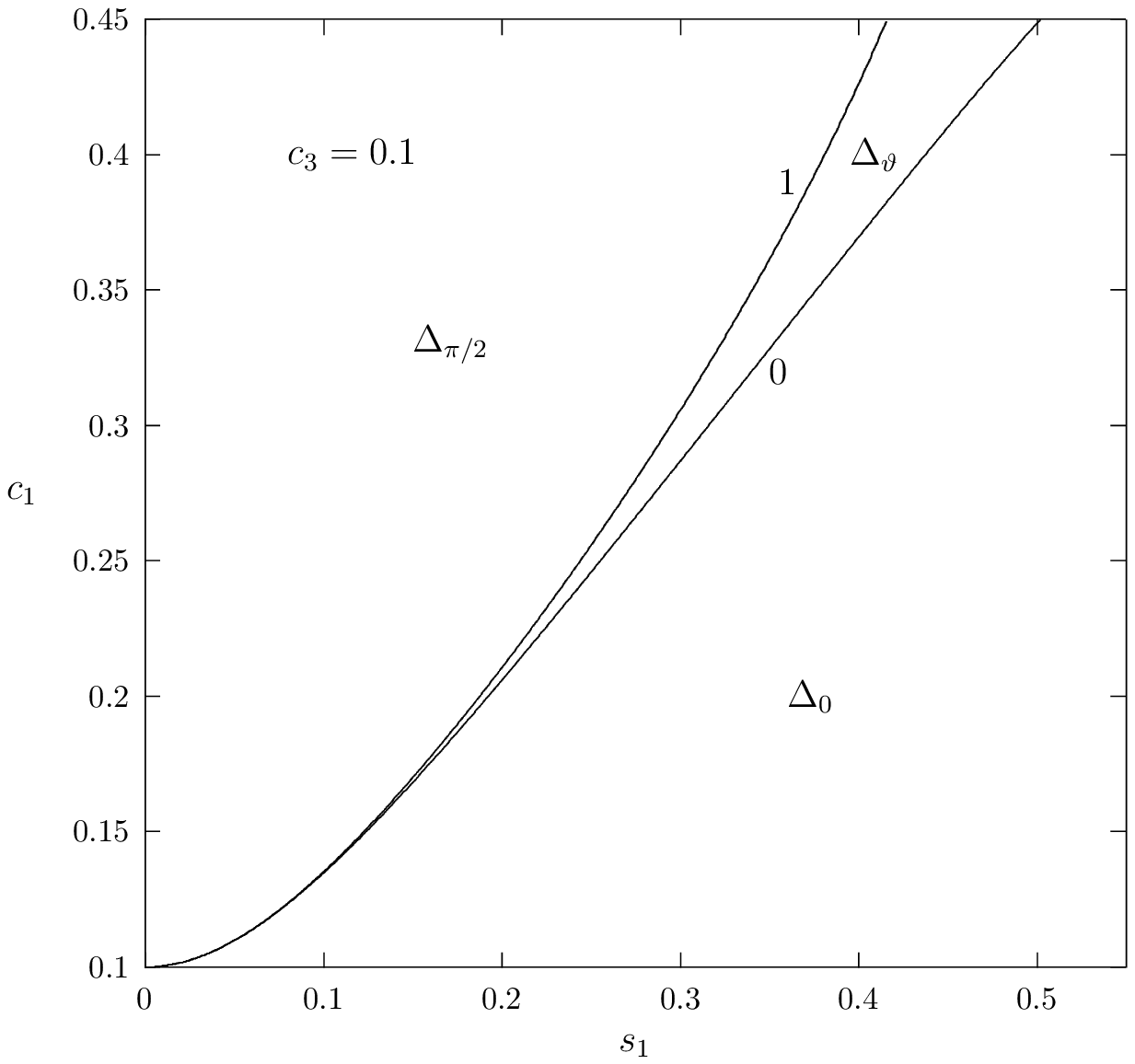,width=5.5cm}
\hspace{0.5cm}
\epsfig{file=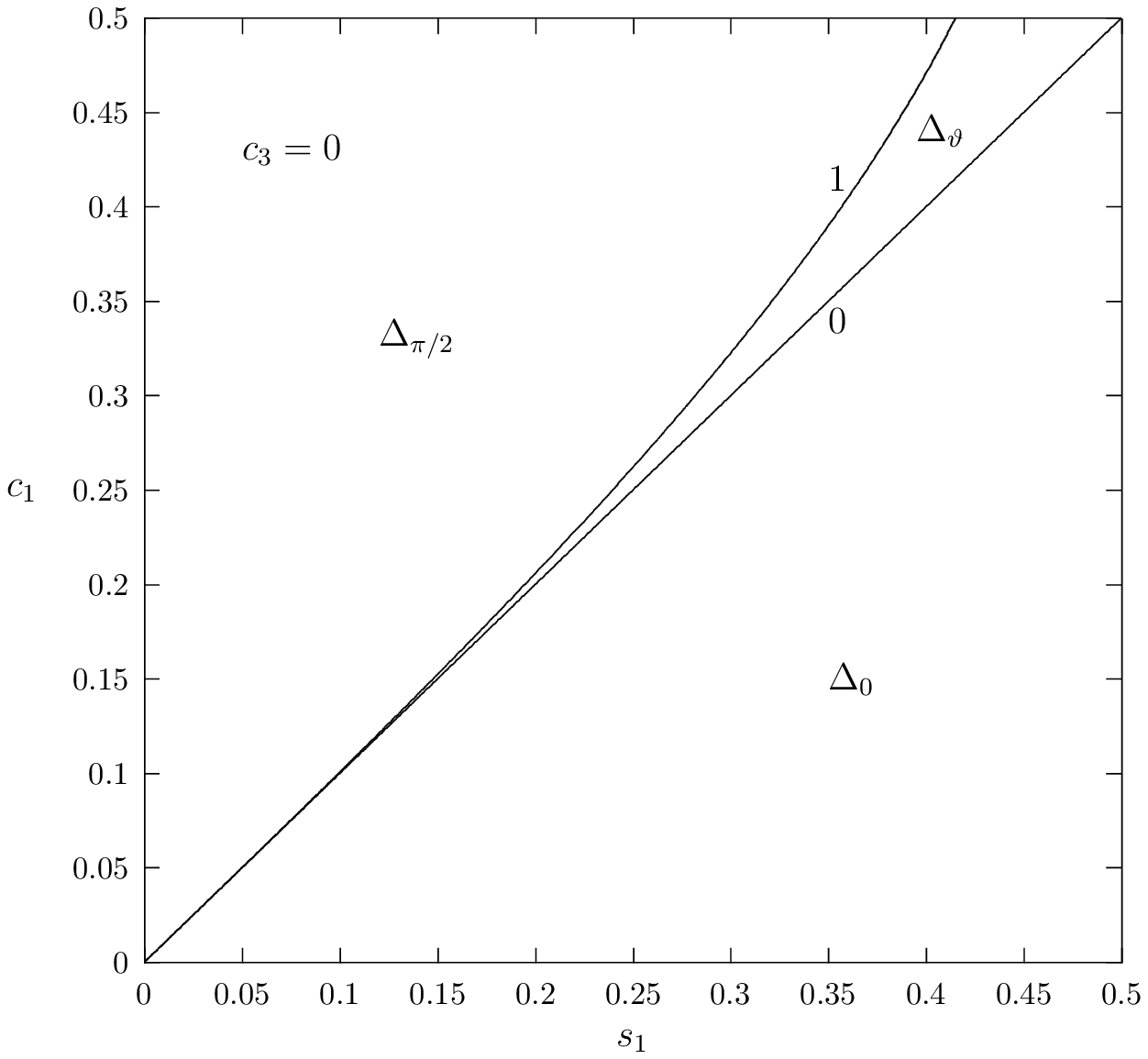,width=5.5cm}
\caption{
Fragments of phase diagrams in the sections $c_3=0.1$ (on the left)
and $c_3=0$ (on the right).
The curves 0 represent the 0-boundaries while the lines 1 are the $\pi/2$-ones
}
\label{fig:pd010}
\end{center}
\end{figure}

Because the phase diagrams in cross-sections are symmetric about the $s_1$
and $c_1$ axes, one may only focus on a quarter of the diagram.
Moreover, as calculations show, the boundaries between phases of one-way deficit lie
in the regions $c_1\geq|c_3|$ therefore it is enough to restrict oneself by the strips
$|c_3|\leq c_1\leq(1-c_3)/2$.
Left part of Fig.~\ref{fig:pd010} shows the subregion by $c_3=0.1$ in detail.
The boundaries 1 and 0 end on the upper edge of cross-section rectangle at the points
with abscissas $s_1=0.416297$ and 0.502469, respectively.
The area of the $\Delta_\vartheta$ segment equals 0.008639 in the absolute units.
Hence the relative area of the whole $\Delta_\vartheta$ region to the area of
cross-section rectangle is 3.5\%.
The similar area for the discord $Q_{\theta^*}$ equals 5.5$\times10^{-5}$\% only.
So, one may ascertain that the region with state-dependent optimal measurement angle
is experimentally accessible for the one-way deficit but not for the discord.

Phase diagram in the cross-section $c_3=0$ is also depicted in Fig.~\ref{fig:pd010}.
The boundary 0 has reached here the right upper corner of the cross-section rectangle,
i.e., the vertex with coordinates $(0.5,0.5)$.
As follow from Eqs.~(\ref{eq:D0_11}) and (\ref{eq:S110}), the 0-boundary is reduced
here to the straight lines $c_1=s_1$.
The endpoint of $\pi/2$-boundary lies at $s_1=0.415037$ on the upper edge of the
cross-section rectangle.
The relative area of the fraction $\Delta_\vartheta$ achieves now 4.2\%.

\begin{figure}[t]
\begin{center}
\epsfig{file=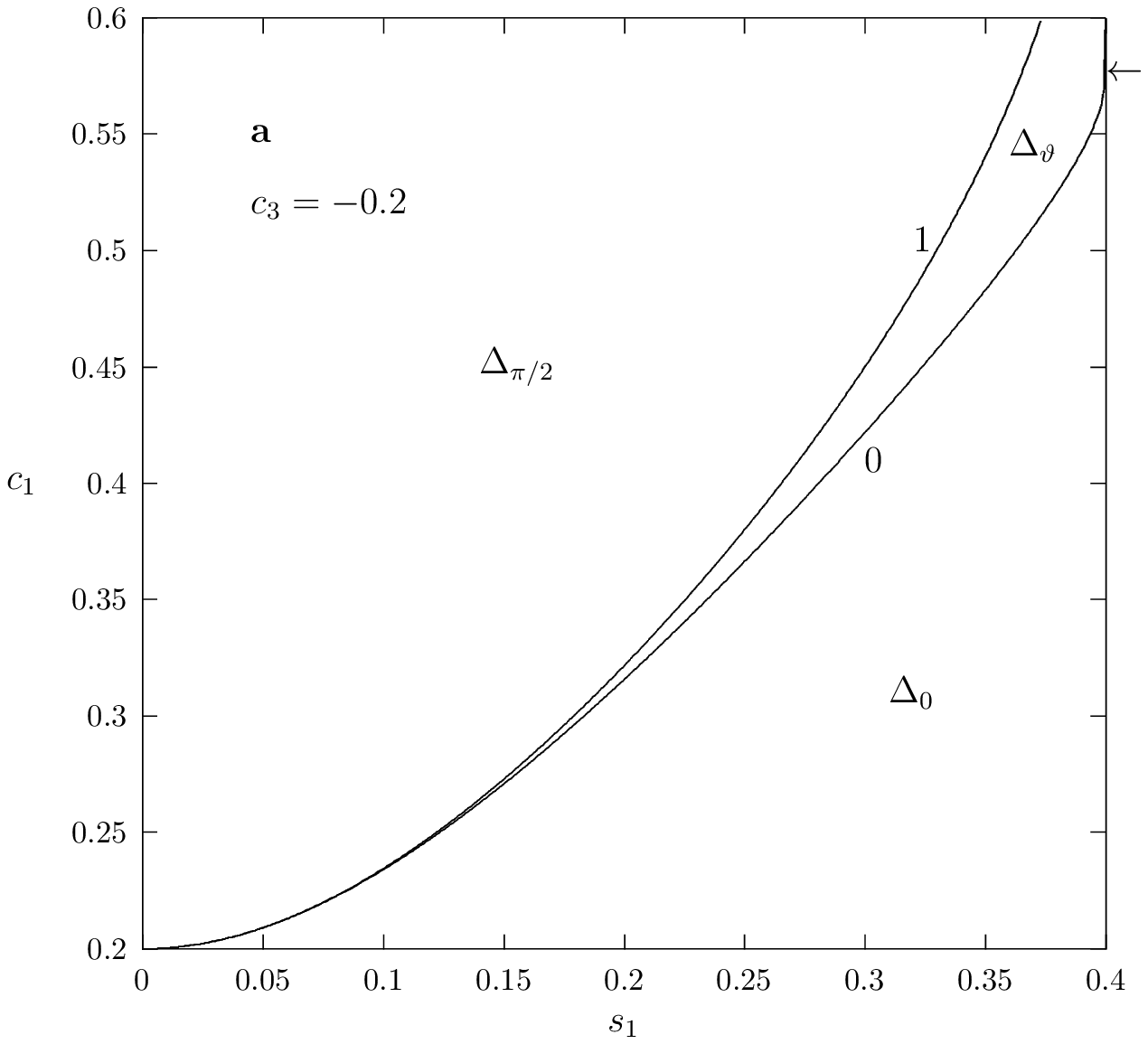,width=5.5cm}
\hspace{0.5cm}
\epsfig{file=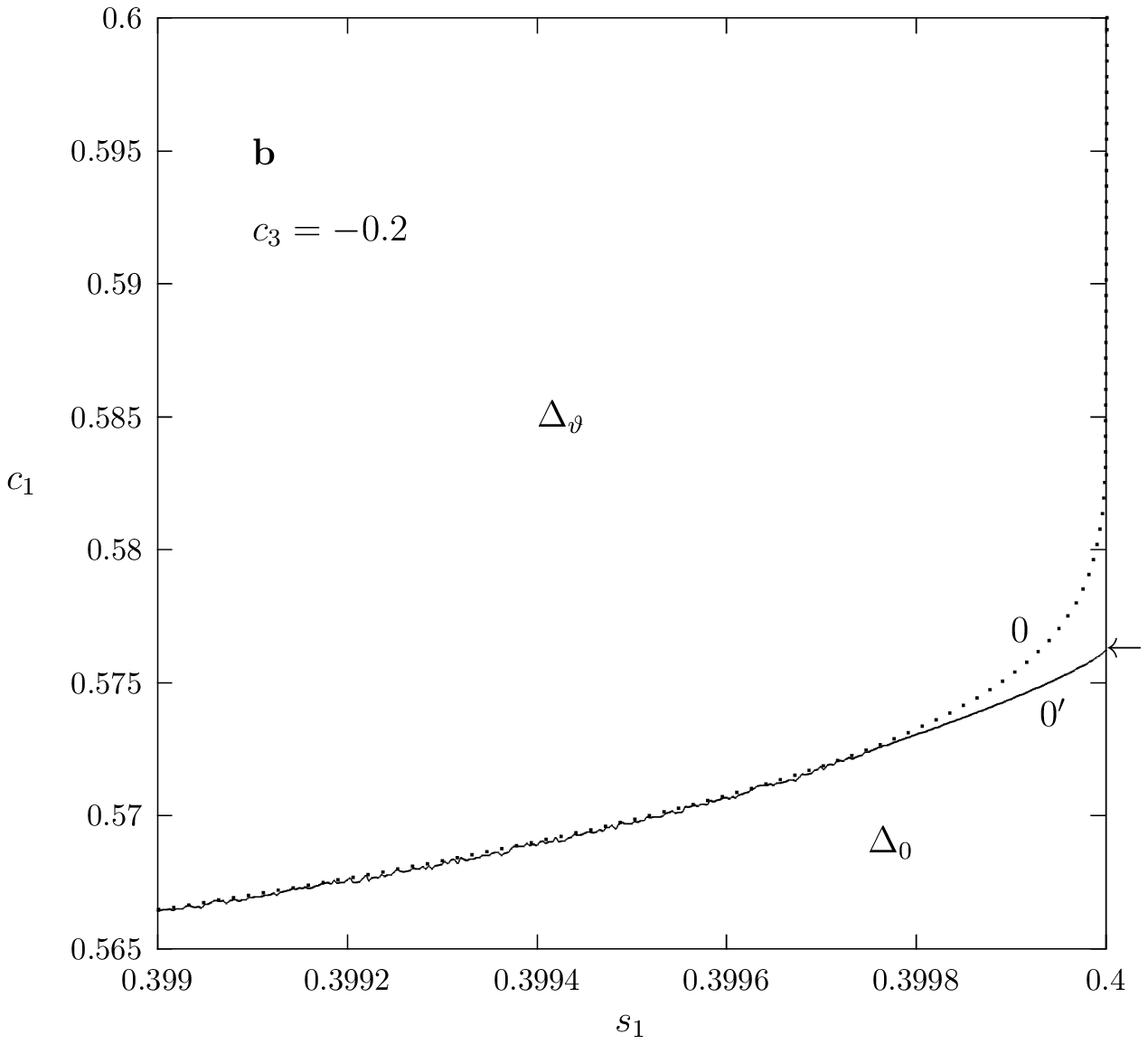,width=5.5cm}
\caption{
Phase diagram by $c_3=-0.2$~(a) and its fragment (b).
The 0-boundary is partly replaced by the $0^\prime$-boundary.
The arrows on the right vertical sides ($s_1=0.4$) mark the point $c_1=0.576208$
which corresponds to the endpoint of the $0^\prime$-boundary
}
\label{fig:pd-02}
\end{center}
\end{figure}

\begin{figure}[t]
\begin{center}
\epsfig{file=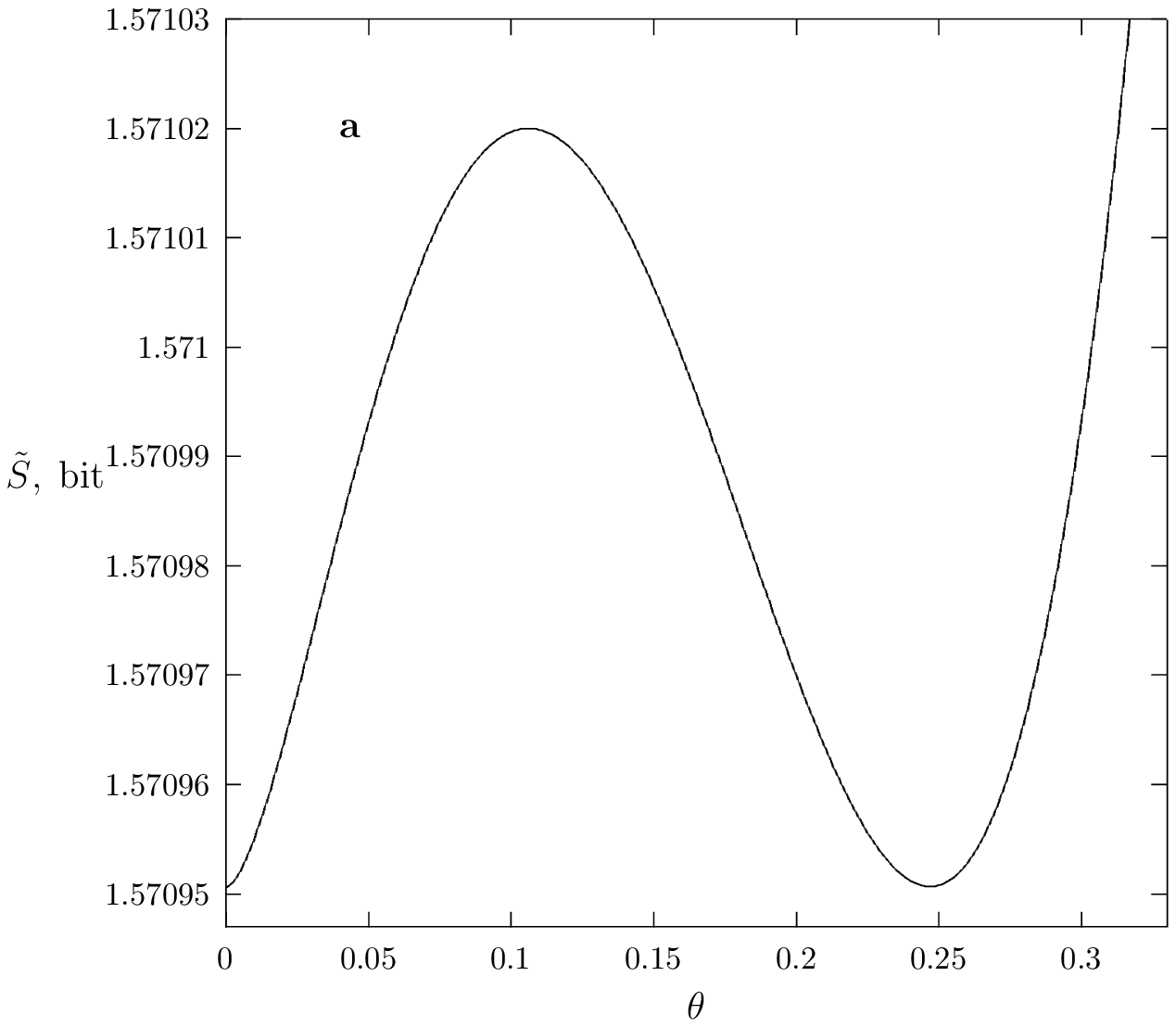,width=5.5cm}
\hspace{0.5cm}
\epsfig{file=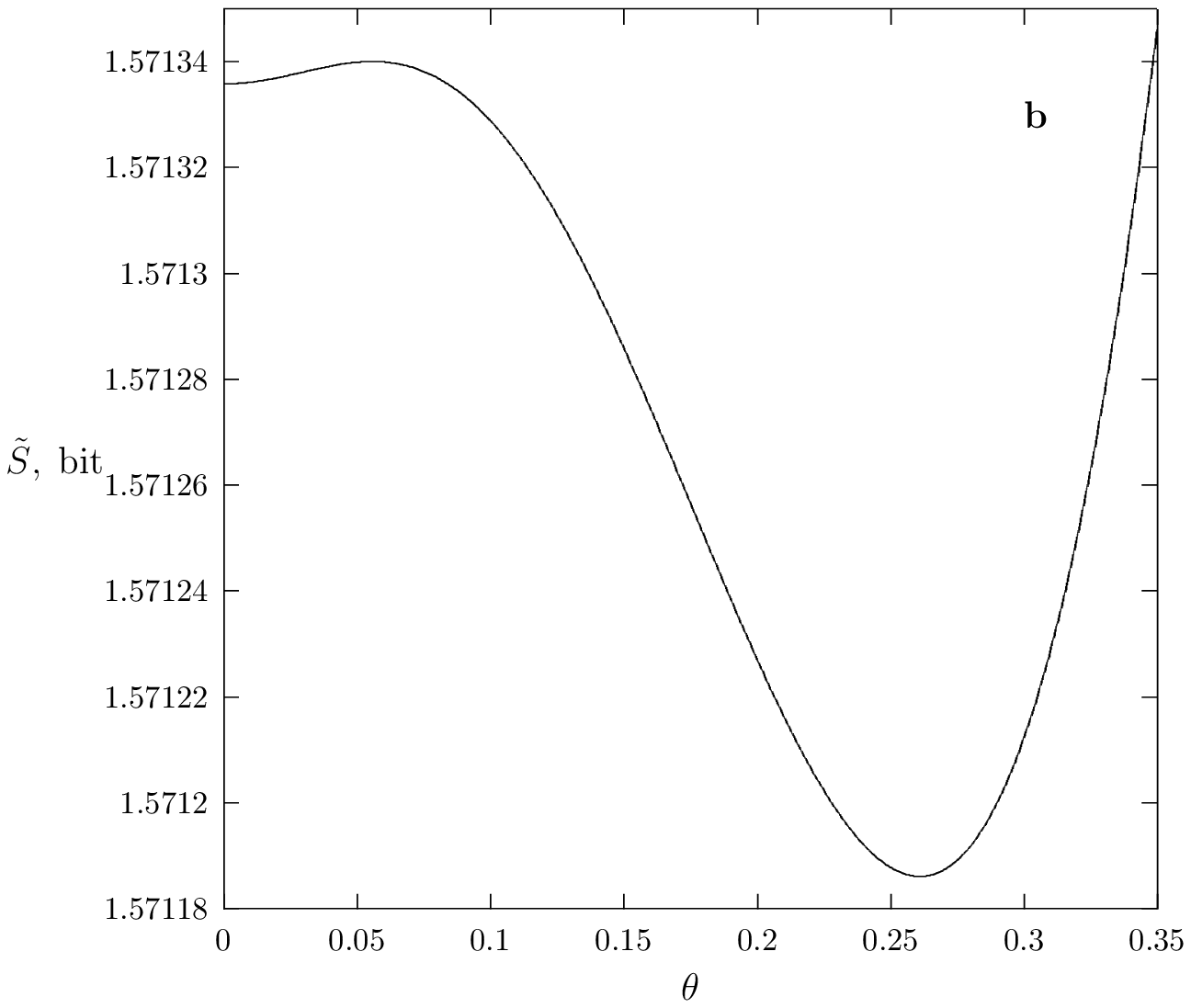,width=5.5cm}
\caption{
Post-measurement entropy $\tilde S$ vs $\theta$ by $c_3=-0.2$, $c_1=0.576208$,
and $s_1=0.4$~(a) and 0.39995~(b).
Both dependencies exhibit the bimodal behavior
}
\label{fig:zs-02}
\end{center}
\end{figure}

The third interval of $c_3$ ranges from $c_3=0$ to $c_3=-0.350302$, i.e.,
up to the point when the $\pi/2$-boundary reaches the vertex of section rectangle
(this corresponds to a segment from the point ``a'' to the point ``b'' in
Fig.~\ref{fig:zzf}).
Characteristic phase diagram in these transverse slices is drawn in
Fig.~\ref{fig:pd-02}.
Near the vertical edge of section rectangle, the 0-boundary is here replaced by
$0^\prime$-boundary, i.e., instead of the appearance of interior minimum via
bifurcation, there is now observed a bimodal behavior of curve $\tilde S(\theta)$
(see Fig.~\ref{fig:zs-02}) that is accompanied with the discontinuous change of
optimal measurement angle on the critical line $0^\prime$.
Figure~\ref{fig:zs-02}a fixes the moment when the interior minimum achieves
the value of post-measurement entropy at the angle $\theta=0$.
Between the lines 0 and 0$^\prime$ the interior minimum is lower than the
minimum at $\theta=0$ as shown in Fig.~\ref{fig:zs-02}b; hence the fraction
$\Delta_\vartheta$ exists here.

\begin{figure}[t]
\begin{center}
\epsfig{file=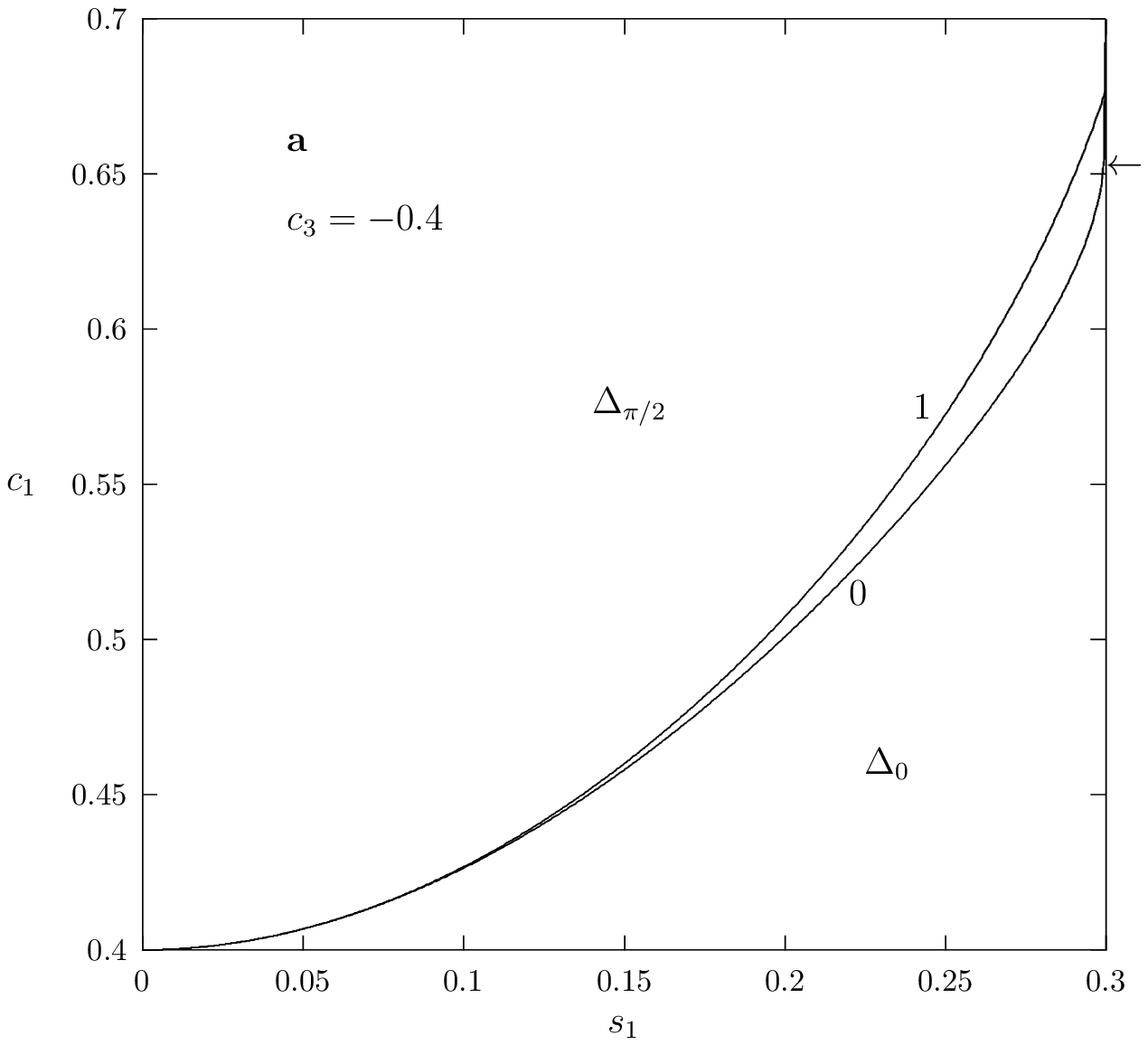,width=5.5cm}
\hspace{0.5cm}
\epsfig{file=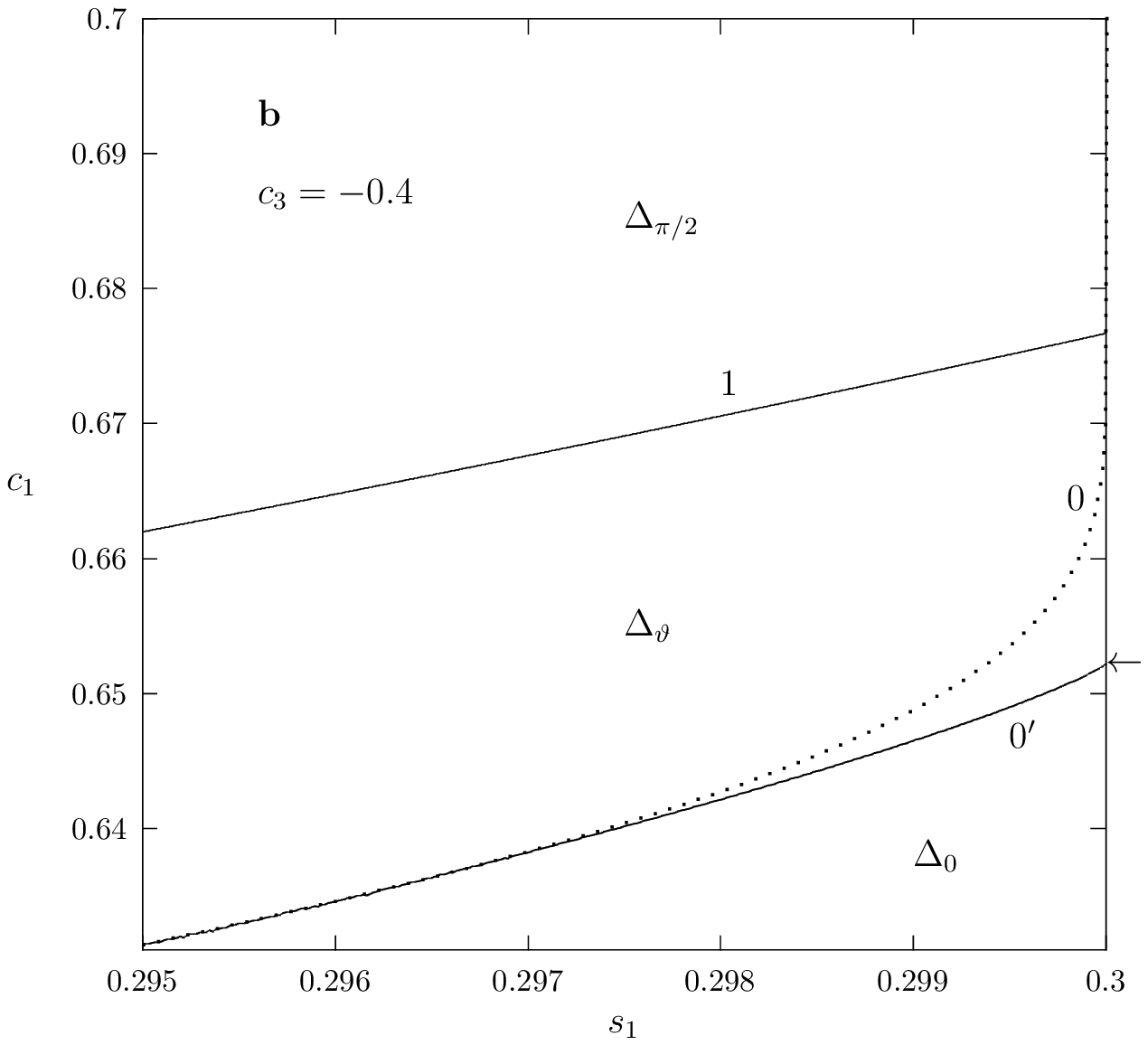,width=5.5cm}
\caption{
Phase diagram in the cross-section $c_3=-0.4$~(a) and its piece
near the vertical edge of cross rectangle~(b).
The arrow marks the point $c_1=0.652165$ of $0^\prime$-boundary on the line $s_1=0.3$
}
\label{fig:pd-04}
\end{center}
\end{figure}

The next interval $-0.538191<c_3<-0.350302$ corresponds to a part of tetrahedron
between the points ``b'' and ``c'' shown in Fig.~\ref{fig:zzf}.
Phase diagram by $c_3=-0.4$ is presented in Fig.~\ref{fig:pd-04}.
As $c_3$ reaches the value of $-0.538191$, the endpoints of curves 1 and
$0^\prime$ meet each other on the side $s_1=(1+c_3)/2$.

\begin{figure}[t]
\begin{center}
\epsfig{file=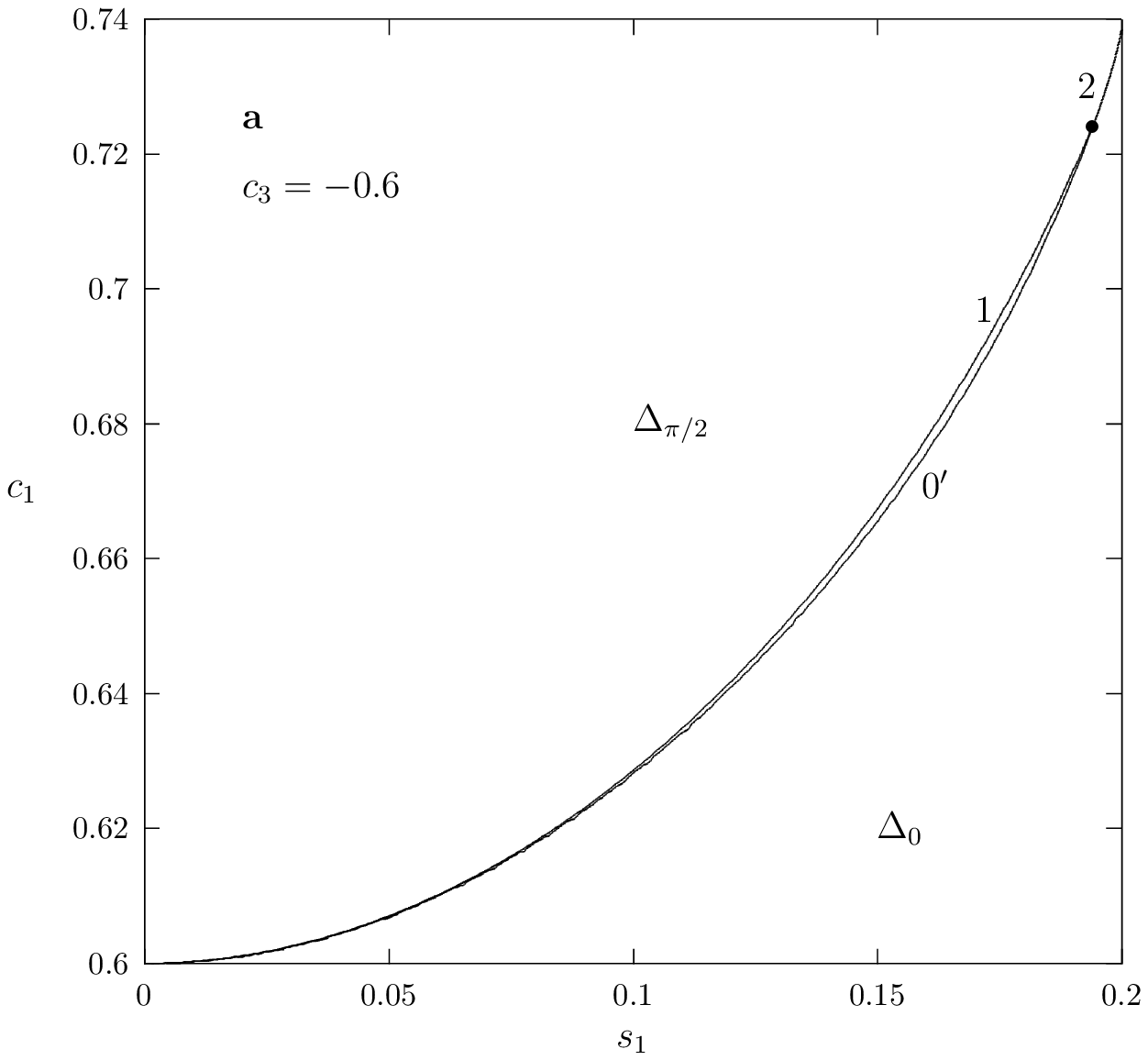,width=5.5cm}
\hspace{0.5cm}
\epsfig{file=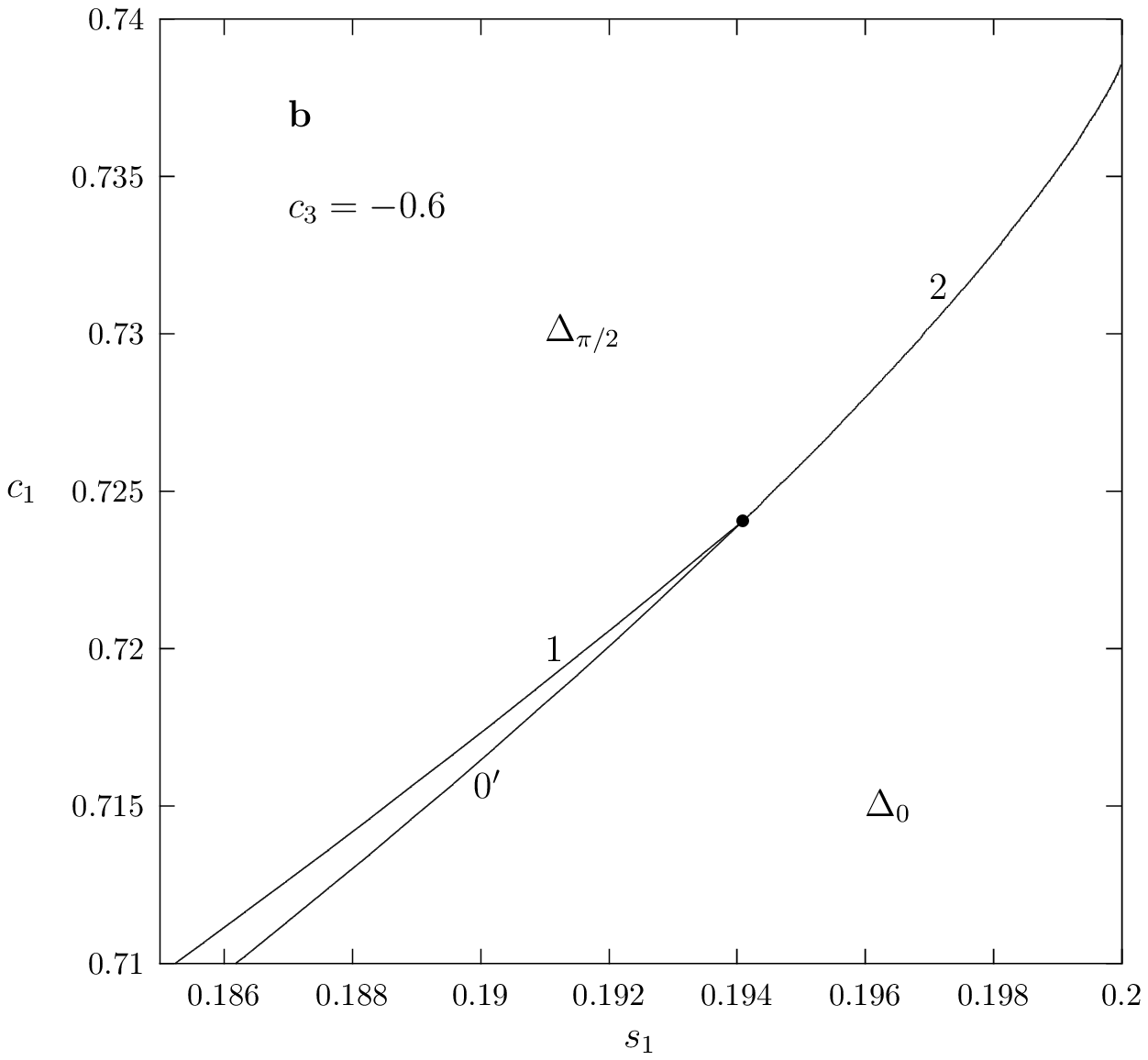,width=5.5cm}
\caption{
Phase diagram by $c_3=-0.6$~(a) and its part~(b).
The lines 1 and $0^\prime$ correspond respectively to the $\pi/2$-
and $0^\prime$-boundaries while the line 2 is the $\Delta_0=\Delta_{\pi/2}$
boundary.
The fraction $\Delta_\vartheta$ lies between the critical lines 1 and $0^\prime$.
Black circle is a triple point
}
\label{fig:pd-06}
\end{center}
\end{figure}

By further decreasing the value of $c_3$, the $\pi/2$- and $0^\prime$-lines 
intersect inside the body $\cal T$ and the boundary
$\Delta_0=\Delta_{\pi/2}$ appears (see Fig.~\ref{fig:pd-06}).
The post-measured entropy curve near the critical line 2 has the interior
{\em maximum}.

In the limit $c_3\to-1$, i.e., on the edge $v_3v_4$ of tetrahedron ${\cal T}$,
the value of parameter $s_1$ vanishes, the region $\Delta_\vartheta$ disappears, and
the quantum state transforms into the  Bell-diagonal one.
On this edge the optimized one-way deficit and discord coincide and are given by
\begin{equation}
   \label{eq:D0Q0v3v4}
   {\rm\Delta}=Q=\frac{1}{2}[(1+c_1)\ln(1+c_1) + (1-c_1)\ln(1-c_1)].
\end{equation}
In the middle of the edge, $c_1=0$, the quantum correlation is zero
while at the vertecies $v_3$ and $v_4$ ($c_1=\pm1$) it is, vice versa, maximal and
equals 1~bit.

\section{Conclusions and outlook}
\label{sect:Concl}
In this paper we have investigated the three-dimensional phase diagram of one-way
quantum deficit for the XXZ family of symmetric X quantum states.
The set of Figs.~\ref{fig:zzf}, \ref{fig:z_xxzb2a}, \ref{fig:pd01}, \ref{fig:pd010},
\ref{fig:pd-02}, \ref{fig:pd-04}, and \ref{fig:pd-06} provides insight into a complete
picture of all typical features in the phase diagram.
It is probable that an application of technologies like holography or virtual reality
would be suitable for the visualization of such 3D diagrams.

It has been established that there are three branches of one-way deficit function and
hence three different subdomains corresponding them.
Two of which, $\Delta_0$ and $\Delta_{\pi/2}$, are characterized by constant
measurement angles --- zero and $\pi/2$, respectively.
At the same time, the third region $\Delta_\vartheta$ is characterized by
non-universal behavior of optimal measurement angle $\vartheta$ because it
continuously varies with the parameters of density matrix.

We have found that three possible regions of one-way deficit can be separated by four
kinds of boundaries:
0 or 0$^\prime$ which divide the $\Delta_0$ and $\Delta_\vartheta$ fractions,
$\pi/2$ that exists between $\Delta_{\pi/2}$ and $\Delta_\vartheta$ phases, and lastly
the $\Delta_0=\Delta_{\pi/2}$ boundary between regions $\Delta_0$ and
$\Delta_{\pi/2}$.
The optimal measurement angle $\vartheta$ is continuous 
when crossing the 0- and $\pi/2$-boundaries, whereas it experiences the jump
$\Delta\vartheta=\pi/2$ by going across the $\Delta_0=\Delta_{\pi/2}$ boundary and
varies in the limits from zero to $\pi/2$ on the critical line 0$^\prime$.

A comparison of behavior of the one-way quantum deficit and discord shows
quantitative and qualitative difference between those measures of quantum
correlation in general.
However they are identical for the Bell-diagonal states, i.e., in the plane $s_1=0$.
Moreover, we have established that both measures of quantum correlation coincide
in the intersection of regions $\Delta_0$ and $Q_0$ ($\Delta_0\cap Q_0$).
(This result is valid for general X states.) 
But quite a difference in other cases allows to say that the named quantities
represent different kinds of quantum correlation.
This situation is similar to the one taking place for the mean value of numbers
and its various types: the arithmetic mean, harmonic mean, and so on.
By this, each kind of mean has its own application.

We have discovered that the region with variable optimal measurement angle is
in several orders larger for the one-way quantum deficit in comparison with the
quantum discord.

It should be noted that all quantum correlations are certain
functions of ordinary statistical correlations (\ref{eq:corr-s1c3})
[see, e.g., Eqs.~(\ref{eq:D0s1c1c3}) or (\ref{eq:Dpi/2s1c1c3})].
This is a sequence of the fact that any quantum correlation is defined by
the system density matrix but its entries are expressed through the statistical
correlation functions.

We have restricted ourselves only by one phase diagram of full atlas.
The work started here should be continued to cover by separate
diagrams the total seven-dimensional space of X-state parameters. 


\vspace{-10mm}
\section*{}
{\bf Acknowledgment}\ I am grateful to Dr.~A.~I.~Zenchuk for his valuable remarks.




\begin{thebibliography}{99}

\bibitem{MBCPV12}
Modi,~K., Brodutch,~A., Cable,~H., Paterek,~T., Vedral,~V.:
The classical-quantum boundary for correlations: discord and related measures.
Rev. Mod. Phys. {\bf 84}, 1655 (2012)

\bibitem{Str15}
Streltsov,~A.:
Quantum correlations beyond entanglement and their role in quantum information
theory.
Springer, Berlin (2015)

\bibitem{ABC16}
Adesso,~G., Bromley,~T.R., Cianciaruso,~M.:
Measures and applications of quantum correlations.
J. Phys. A: Math. Theor. {\bf 49}, 473001 (2016)

\bibitem{FPA17}
Lectures on general quantum correlations and their applications.
Eds: Fanchini,~F.F., Soares-Pinto,~D.O., Adesso,~G.
Springer, Berlin (2017)

\bibitem{BDSRSS18}
Bera,~A., Das,~T., Sadhukhan,~D., Roy,~S.S., Sen(De),~A., Sen,~U.:
Quantum discord and its allies: a review of recent progress.
Rep. Prog. Phys. {\bf 81}, 024001 (2018)

\bibitem{OHHH02}
Oppenheim,~J., Horodecki,~M., Horodecki,~P., Horodecki,~R.:
Thermodynamical approach to quantifying quantum correlations.
Phys. Rev. Lett. {\bf 89}, 180402 (2002)

\bibitem{HHHHOSS02}
Horodecki,~M., Horodecki,~K., Horodecki,~P., Horodecki,~R., Oppenheim,~J., Sen(De), A., Sen, U.:
Local information as a resource in distributed quantum systems.
Phys. Rev. Lett. {\bf 90}, 100402 (2003)

\bibitem{HHHOSSS05}
Horodecki,~M., Horodecki,~P., Horodecki,~R., Oppenheim,~J., Sen(De), A., Sen, U., Synak-Radtke, B.:
Local versus nonlocal information in quantum-information theory: Formalism and phenomena.
Phys. Rev. A {\bf 71}, 062307 (2005)

\bibitem{Z03}
Zurek,~W.H.:
Quantum discord and Maxwell's demons.
Phys. Rev. A {\bf 67}, 012320 (2003)

\bibitem{YF16}
Ye,~B.-L., Fei,~S.-M.:
A note on one-way quantum deficit and quantum discord.
Quantum Inf. Process. {\bf 15}, 279 (2016)

\bibitem{CRC10}
Ciliberti,~L., Rossignoli,~R., Canosa,~N.:
Quantum discord in finite $XY$ chains.
Phys. Rev. A {\bf 82}, 042316 (2010)

\bibitem{VR12}
Vinjanampathy,~S., Rau,~A.R.P.:
Quantum discord for qubit-qudit systems.
J. Phys. A: Math. Theor. {\bf 45}, 095303 (2012)

\bibitem{H13}
Huang,~Y.:
Quantum discord for two-qubit $X$ states: Analytical formula with very small worst-case error.
Phys. Rev. A {\bf 88}, 014302 (2013)

\bibitem{YWF16}
Ye,~B.-L., Wang,~Y.-K., Fei,~S.-M.:
One-way quantum deficit and decoherence for two-qubit $X$ states.
Int. J. Theor. Phys. {\bf 55}, 2237 (2016)

\bibitem{Y17}
Yurischev,~M.A.:
Extremal properties of conditional entropy and quantum discord for XXZ, symmetric quantum states.
Quantum Inf. Process. {\bf 16}:249 (2017)

\bibitem{BM12}
Brodutch,~A., Modi,~K.:
Criteria for measures of quantum correlations.
Quantum Inf. Comput. {\bf 12}, 0721 (2012)

\bibitem{Y14}
Yurischev,~M.A.:
Quantum discord for general X and CS states: a piecewise-analytical-numerical formula.
ArXiv:1404.5735v1 [quant-ph]

\bibitem{Y14a}
Yurishchev,~M.A.:
NMR dynamics of quantum discord for spin-carrying gas molecules in a closed nanopore.
J. Exp. Theor. Phys. {\bf 119}, 828 (2014),
arXiv:1503.03316v1~[quant-ph]

\bibitem{P87}
Postnikov,~M.M.:
Lectures in geometry. Semester~III. 
Smooth manifolds.
Nauka, Moscow (1987), lecture~6 [in Russian]

\bibitem{B82}
Baxter,~R.J.:
Exactly solved models in statistical mechanics.
Academic, London (1982)

\bibitem{F14}
Fendley,~P.:
Modern statistical mechanics.
The university of Virginia (2014)

\bibitem{BMNM04}
Barbieri,~M., De~Martini,~F., Di~Nepi,~G., Mataloni,~P.:
Generation and characterization of Werner states and maximally entangled mixed states
by a universal source of entanglement.
Phys. Rev. Lett. {\bf 92}, 177901 (2004)

\bibitem{MMH17}
Mendonca,~P.E.M.F, Marchiolli,~M.A., Hedemann,~S.R.:
Maximally entangled mixed states for qubit-qutrit systems.
Phys. Rev. A {\bf 95}, 022324 (2017)

\bibitem{IH00}
Ishizaka,~S., Hiroshima,~T.:
Maximally entangled mixed states under nonlocal unitary operations in two qubits.
Phys. Rev. A {\bf 62}, 022310 (2000)

\bibitem{HI00}
Hiroshima,~T., Ishizaka,~S.:
Local and nonlocal properties of Werner states.
Phys. Rev. A {\bf 62}, 044302 (2000)

\bibitem{PABJWK04}
Peters,~N.A., Altepeter,~J.B., Branning,~D., Jeffrey,~E.R., Wei,~T.-C., Kwiat,~P.G.:
Maximally entangled mixed states: creation and concentration.
Phys. Rev. Lett. {\bf 92}, 133601 (2004);
Erratum in: Phys. Rev. Lett. {\bf 96}, 159901 (2006)

\bibitem{APVW07}
Aiello,~A., Puentes,~G., Voigt,~D., Woerdman,~J.P.:
Maximally entangled mixed-state generation via local operations.
Phys. Rev. A {\bf 75}, 062118 (2007)

\bibitem{KHJP10}
Kim,~H., Hwang,~M.-R., Jung,~E., Park,~D.K.:
Difficulties in analytic computation for relative entropy of entanglement.
Phys. Rev. A {\bf 81}, 052325 (2010)

\bibitem{GGZ11}
Galve,~F., Giorgi,~G.L., Zambrini,~R.:
Maximally discordant mixed states of two qubits.
Phys. Rev. A {\bf 83}, 012102 (2011);
Erratum in: Phys. Rev. A {\bf 83}, 069905 (2011)

\bibitem{MHR15}
Maldonado-Trapp,~A., Hu,~A., Roa,~L.:
Analytical solutions and criteria for the quantum discord of two-qubit X-states.
Quantum Inf. Process. {\bf 14}, 1947 (2015)

\bibitem{SXL13}
Shao,~L.-H., Xi,~Z.-J., Li,~Y.M.:
Remark on the one-way quantum deficit for general two-qubit states.
Commun. Theor. Phys. {\bf 59}, 285 (2013)

\bibitem{Y15}
Yurischev,~M.A.:
On the quantum discord of general X states.
Quantum Inf. Process. {\bf 14}, 3399 (2015)

\bibitem{Y18}
Yurischev,~M.A.:
Bimodal behavior of post-measured entropy and one-way quantum deficit
for two-qubit X states.
Quantum Inf. Process. {\bf 17}:6 (2018)

\bibitem{LL_StPh}
Landau,~L.D., Lifshitz,~E.M.:
Statistical physics. Part~1.
Fizmatlit, Moscow (2005) [in Russian],
Pergamon, Oxford (1980) [in English]

\end{thebibliography}
\end{document}